\documentclass[twocolumn]{aastex631}

\usepackage{color}
\usepackage{subfigure}
\usepackage{booktabs}
\usepackage{longtable}
\usepackage{soul}
\usepackage{bigstrut}
\usepackage{graphicx}
\usepackage{amsmath}

\usepackage{mathrsfs}

\graphicspath{{./}{figures/}}

\revised{\today}

\shorttitle{Tidal Peeling Events}
\shortauthors{Xin et al.}

\begin{document}

\title{Tidal Peeling Events: low-eccentricity tidal disruption of a star by a stellar-mass black hole}

\correspondingauthor{Chengcheng Xin}
\email{cx2204@columbia.edu}

\author[0000-0003-3106-8182]{Chengcheng Xin}
\affil{Columbia University, Department of Astronomy, 550 West 120th Street, New York, NY, 10027, USA}

\author{Zolt{\'{a}}n~Haiman}
\affil{Columbia University, Department of Astronomy, 550 West 120th Street, New York, NY, 10027, USA}
\affil{Columbia University, Department of Physics, 550 West 120th Street, New York, NY, 10027, USA}

\author[0000-0002-3635-5677]{Rosalba Perna}
\affil{Department of Physics and Astronomy, Stony Brook University, Stony Brook, NY 11794-3800, USA}
\affil{Center for Computational Astrophysics, Flatiron Institute, New York, NY 10010, USA}

\author{Yihan Wang}
\affil{Nevada Center for Astrophysics, University of Nevada, 4505 S. Maryland Pkwy., Las Vegas, NV 89154-4002, USA}
\affil{Department of Physics and Astronomy, University of Nevada, 4505 S. Maryland Pkwy., Las Vegas, NV 89154-4002, USA}

\author{Taeho Ryu}
\affil{Max Planck Institute for Astrophysics, Karl-Schwarzschild-Strasse 1, 85748 Garching, Germany}
\affil{Physics and Astronomy Department, Johns Hopkins University, Baltimore, MD 21218, USA}

\begin{abstract}
 Close encounters between stellar-mass black holes (BHs) and stars occur frequently in dense star clusters and in the disks of active galactic nuclei (AGNs). Recent studies have shown that in highly eccentric close encounters, the star can be tidally disrupted by the BH (micro-tidal disruption event, or micro-TDE), resulting in rapid mass accretion and possibly bright electromagnetic signatures. Here we consider a scenario in which the star might approach the stellar-mass BH in a gradual, nearly circular inspiral, under the influence of dynamical friction on a circum-binary gas disk or three-body interactions in a star cluster. We perform hydro-dynamical simulations of this scenario using the smoothed particle hydrodynamics code {\sc PHANTOM}. We find that the mass of the star is slowly stripped away by the BH. We call this gradual tidal disruption a “tidal-peeling event”, or a TPE. Depending on the initial distance and eccentricity of the encounter, TPEs might exhibit significant accretion rates and orbital evolution distinct from those of a typical (eccentric) micro-TDE. 
\end{abstract}

\section{Introduction}
\label{sec:intro}

Stars and their compact remnants, which include stellar-mass black holes (BHs), are expected to be abundant in dense stellar clusters of all kinds \citep{Mackey2007,Strader2012}, and they can also be found in the disks of Active Galactic Nuclei (AGNs). Dynamical interactions between compact objects and stars in clusters are frequently expected \citep{Rodriguez2016,Kremer2018}. As a result, stars in a cluster will inevitably undergo close encounters with stellar-mass BHs. These close encounters between stars and BHs, which are of particular interest here, can lead to binary formation or to tidal disruption of the star by the BH (the so-called micro-TDEs, \citealt{Perets2016}).

Stars and stellar-mass BHs  found in an AGN disk are likely the result of two mechanisms: 
\textit{ (i)} Capture from the nuclear star cluster \citep{Artymowicz1993}, which consists  mostly of massive stars (e.g. O- and B-type stars with masses $\gtrsim$ 2-15$M_{\odot}$). These stars' orbits will eventually align with the AGN disk after a number of crossings of the disk \citep{Yang2020}.
\textit{(ii)} In-situ formation: Gravitational instabilities in the outer parts of the disk trigger star formation \citep{Goodman2003,Dittmann2020}, and those stars, as well as their remnant compact objects, remain embedded in the disk. The unusual disk environment causes stars to accrete and grow in mass \citep{Cantiello2021,Jermyn2021}, which makes BH remnants a common outcome upon their death.
Once trapped in the AGN disk, BHs can go through radial migration and undergo close encounters with stars or compact objects \citep[e.g.,][]{Tagawa2020}. Therefore, micro-TDEs can also occur in AGN disks, in addition to the stellar cluster environment.

Micro-TDEs are expected to be ultra-luminous events, and their expected accretion rates and 
electromagnetic (EM) features have recently begun to be investigated in more detail via 
smooth particle hydrodynamical (SPH) simulations
(\citealt{Lopez2019,Kremer2021,Wang2021,Kremer2022,Ryu2022a}) and moving-mesh \citep{Ryu2023}. 
Existing studies have performed numerical experiments to investigate nearly parabolic encounters with eccentricity $e\sim 1$. \citet{Kremer2022} recently presented a variety of hydrodynamical simulations of the typical micro-TDE with parabolic orbits to show that stars {\it in vacuum} can experience different degrees of tidal disruption depending on pericenter distance and stellar mass, while the peak luminosity of the EM emission might be super-Eddington when pericenter distance is within $\sim2R_t$, where $R_t=(M_{\rm BH}/M_{\rm s})^{1/3} R_{\rm s}$ is the order-of-magnitude estimate of the tidal radius for a star with mass $M_{\rm s}$ and radius $R_{\rm s}$ disrupted by a BH with mass $M_{\rm BH}$.

On the other hand, low-eccentricity micro-TDEs in compact orbits are of particular interest in this paper for the following reasons. 
First, observational work has suggested that binaries in clusters have lower eccentricity as they become more compact \citep{Meibom2005,Hwang2022}. 3D hydro-simulations by \cite{Ryu2023} further suggest that three-body interactions in clusters such as encounters between binary stars and stellar-mass BHs can also lead to eventual close interactions between one star in the original binary and the BH, where, in some cases, a low-eccentricity micro-TDE in a close orbit can form if the star becomes bound to the BH. 
Additionally, star-BH binaries in an AGN disk can become tightly bound due to external torques exerted by the dynamical friction of the AGN disk gas. 
Hydrodynamical simulations have shown that a circumbinary disk tends to shrink the orbit of the binary within an AGN disk \citep{Li2021,Kaaz2021,Li2022} and drive it to low eccentricity, either $e\rightarrow0$ or $e\rightarrow 0.45$, depending on the initial value \citep{Munoz2019a,DOrazio2021,Zrake2021}.

Unlike the abrupt disruption that the star experiences in a parabolic TDE or micro-TDE, lower-eccentricity micro-TDEs gradually strip mass from the star, typically over many orbital times, analogous to the extreme-mass-ratio inspiral of a white dwarf (WD) and an intermediate mass BH, in which the WD loses mass periodically during the inspiral \citep{Zalamea2010,Chen2022}.
% where some stellar material is accreted onto the companion BH;
We call this ``tidal-peeling event" (TPE) in this paper. 

In this paper, we numerically model the general case of TPEs with SPH simulations using \texttt{PHANTOM}, without  including the low-density background gas such as the AGN disk. We focus on exploring the BH mass accretion rate and orbital evolution in TPEs under different assumptions for the initial mass of the star, eccentricity and pericenter distance of the encounter. 
% We fix the BH mass in all the simulation models at $M_{\rm BH}=10M_{\odot}$. 

We organize this paper as follows. We describe our simulation models, analysis method and a resolution study in \S~\ref{sec:method}, \ref{sec:analysis} and \ref{sec:resolution}, respectively. In \S~\ref{sec:morphology}, we show the morphological evolution of the TPEs. Section \S~\ref{sec:em_features} illustrates our prediction for the EM signatures of TPEs, based on the computation of the BH mass accretion rates, stellar mass loss via tidal interactions and the orbit evolution of the remnant. In \S~\ref{sec:massive_stars}, we explore the effect of having more massive stars undergoing TPEs. Finally, we discuss some implications of our results in \S~\ref{sec:discussion}, and we summarize our conclusions in \S~\ref{sec:summary}.

% \noteTH{The Introduction is written in the context of encounters in AGN disks whereas we simulate encounters in a vacuum. I think we may want to justify why we do not include a gas medium in our simulations. }

\section{Simulation Methods} \label{sec:method}
We perform SPH simulations of TPEs of stars by a $10M_{\odot}$ BH using \texttt{PHANTOM} \citep{Price2018}. We run simulations for (4 stellar masses) $\times$  (4 eccentricities) $\times$ (6 penetration factors) $=96$ models in total, where the penetration factor $\beta$ is defined as the ratio between the tidal radius and the pericenter distance, or $R_t/r_p$. 
We consider main-sequence (MS) stars with four different masses, $M_{\rm s}=$ 1, 5, 10 or 15 $M_{\odot}$, and investigate the dependence of the initial eccentricities of outcomes by considering  $e_0=$ 0.0, 0.2, 0.4 and 0.6. We begin all simulations by placing the star at the apocenter of the orbit. 
Finally, we consider the following penetration factors $\beta=R_t/r_p$ = 1, 0.67, 0.5, 0.4, 0.33 and 0.25, which corresponds to the pericenter distances $r_p=$ 1, 1.5, 2, 2.5, 3 and 4 times the tidal radii.  For simplicity, we introduce the letter $\mathcal{M}(M_{\rm s},e_0,\beta)$ to denote any specific model, where $M_{\rm s}$ is given in units of $M_{\odot}$. We fix the BH mass in all the simulation models at $M_{\rm BH}=10M_{\odot}$. 

We first use the 1D stellar evolution code \texttt{MESA} \citep{Paxton+2019} to generate the profile of each MS star with the core H fraction of 0.5, where we assume solar abundances for composition, hydrogen and metal mass fractions $X=0.74$ and $Z=0.02$ respectively (helium mass fraction $Y=1-X-Z$), and mean molecular weight $\mu\sim0.59$ (fully ionized gas). For the stellar masses that we consider, \texttt{MESA} uses the OPAL and HELM table for the equation of state \citep{Paxton+2019}, which we adopt in the TPE simulations.
We then take the density and internal energy profile of \texttt{MESA} MS stars to start the simulations in \texttt{PHANTOM}. 
We first map the 1D \texttt{MESA} model onto our 3D SPH grid and relax it for a few stellar dynamical times ($t_{\rm dyn}=\sqrt{R_{s}^{3}/GM_{s}}$) until it reaches hydrostatic equilibrium. 
$t_{\rm dyn}$ is typically 1 to a few hours depending on the mass and radius of the star. 

In the TPE simulations with \texttt{PHANTOM}, we use artificial viscosity varying between ${a^{\rm AV}}_{\rm min}=0.1$ to ${a^{\rm AV}}_{\rm max}=1$. This is the typical range for ${a^{\rm AV}}$ to evolve, which contributes to shock capture \citep[e.g.][]{Coughlin2017}.
We adopt an equation of state that includes radiation pressure assuming instantaneous local thermodynamic equilibrium. This assumption is valid because the gas in our simulations is expected to be optically thick. 
% For the accretion radius of the BH, or the radius of the sink particle, we adopt $r_{\rm acc}=100r_g$, where $r_g=GM_{\rm BH}/c^2$ is the gravitational radius. This is the smallest radius that exceeds the size of the SPH particle, or the minimum radius at which accretion can be resolved.
% The particles are removed from the simulation once accreted by the BH; the removed mass is added to the mass of the sink particle. Finally, we employ $10^5$ SPH particles in each simulation, which is justified in \S~\ref{sec:resolution}, and each simulation uses up to 6,000 CPU hours on processor Intel Xeon Gold 6226 2.9 Ghz.
% 
We employ $10^5$ SPH particles in each simulation, which is justified in \S~\ref{sec:resolution}, and each simulation uses up to 6,000 CPU hours on processor Intel Xeon Gold 6226 2.9 Ghz. For this resolution, the smallest spatial scale within which accretion can be resolved is $r_{\rm acc}=100r_g$, where $r_g=GM_{\rm BH}/c^2$. If a SPH particle falls within the ``accretion`` radius, it is accreted onto the BH. The particles are removed from the simulation once accreted by the BH; the removed mass is added to the mass of the sink particle. 

% \noteTH{Could you add the criteria for accretion? If you follow the accretion critera already implemented in the code without making any modifications, I think you can simply refer to their code paper. } \cx{we didn't modify the accretion criteria, but I think our description is simple enough.. }

\section{Analysis} \label{sec:analysis}
In this study, we focus on some key physical quantities, such as the amount of mass lost in TPEs and the accretion rate, directly measured from our simulation output. Also, we investigate their dependence on different initial conditions -- the mass of the star ($M_s$), 
% \addTH{semimajor axis} (SMA)
 the initial eccentricity ($e_0$), and the penetration parameter ($\beta$) that is inversely proportional to the initial pericenter distance.

First, we measure the mass accretion onto the BH, $M_{\rm acc}$, by evaluating the mass accreted onto the sink particle representing the BH. The BH accretion rate $\dot{M}_{\rm BH}$ is computed as the finite difference of $M_{\rm acc}$ divided by the time difference ($\sim 0.4$ hours) between two adjacent outputs of the simulation. 
% \noteTH{Two adjacent time steps or two adjacent output files?}

In a TPE, the star's mass is slowly stripped by the BH, which leads to the star being partially or totally disrupted. In past studies of TDEs or micro-TDEs using numerical simulations \citep[e.g.,][]{Mainetti2017,Kremer2022}, the mass bound to the star or BH is usually computed using an iterative process described in \citet{Lombardi2006}. However, since the iteration evaluates the specific binding energy of each particle, including a gravitational potential term, it assumes spherical geometry for the remnant, which is not always applicable in our TPE simulations, see Fig.~\ref{fig:morphology4} for example. Additionally, in some TPEs, the remnant is not isolated as it is connected with debris, for which the iterative process can lead to inaccurate identification of the remnant. Alternatively, we define the mass of the stellar remnant ($M_{\rm rem}$) as the total mass of particles within the initial radius of the star (measured from the densest point in the star). 

In addition to the stellar material lost to $M_{\rm acc}$, the star can also lose mass to the surroundings when the stellar material is unbound during the disruptions. We measure the fraction of total mass removed from the star, $f_{\rm rm}$. The mass removed consists of mass accreted by the BH ($M_{\rm acc}$) and mass ejected (total stellar mass minus remnant mass; $M_{\rm s}$-$M_{\rm rem}$). Note that the mass removed from the star includes the mass unbound to the remnant, but bound and not yet accreted by the BH. So $f_{\rm rm}$=($M_{\rm s}$ - $M_{\rm rem}$ + $M_{\rm acc}$)/$M_s$.

The orbital features of the stellar remnant can be described by the evolution of the orbital separation ($r$), semi-major axis  (SMA; $a$) and eccentricity ($e$) over time. We define $r$ to be the distance between the particle of the highest density in the stellar remnant, typically at the core of the star (small deviation can happen due to any oscillation in the star during the disruption), and the position of the sink particle (BH). 
% The specific angular momentum is $\vec{l}=\vec{r}\times \vec{v}$, where $\vec{r}$ and $\vec{v}$ are the relative position and velocity of the SPH particle with respect to the BH. The semi-major axis and eccentricity are given in eqs. \ref{eq:sma} and \ref{eq:ecc}. 
The SMA and the eccentricity are calculated using the specific energy and specific angular momentum of the binary, adapted from the calculation in \cite{Munoz2019a}, where the equation of motion of the binary are evaluated with the external gravitational and accretion forces.
In \S~\ref{sec:em_features}, we evaluate the evolution of  $a$ and $e$, as well as their change per each orbit around the BH. 
% \begin{equation} \label{eq:sma}
%     a = \frac{|\vec{l}|^2}{\mu} \frac{1}{1-e^2} ; \mu\approx G M_{\rm BH}
% \end{equation}
% \begin{equation}\label{eq:ecc}
%     \vec{e} = \frac{\vec{v} \times \vec{l}}{\mu} - \frac{\vec{r}}{|\vec{r}|} ; e = |\vec{e}|.
% \end{equation}

\section{Resolution tests for initial stellar profile} \label{sec:resolution}
\begin{figure}
    \centering
    \includegraphics[width=\columnwidth]{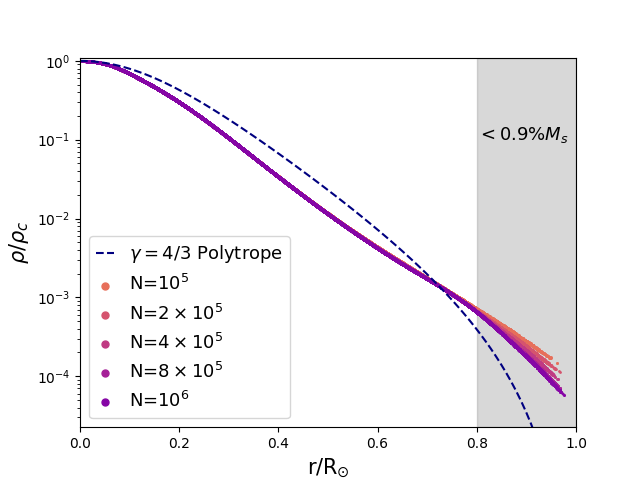}
    \caption{The radial density profile of a fully relaxed $1M_{\odot}$ star in \texttt{PHANTOM}, using $N=10^5, 2\times10^5, 4\times10^5, 8\times10^5, 10^6$ SPH particles. The density is normalized to the core density $\rho_{\rm c}$. 
    % For a stellar mass $\gtrsim1M_{\odot}$, the density profile is more consistent with that of a $\gamma=4/3$ polytrope (dashed navy line). 
    Different resolutions yield converging initial density profiles for the star, despite a small surface layer (R$>0.8R_{\odot}$; gray region), containing $\lesssim$0.9\% of stellar mass. This justifies our choice to use $N=10^5$ particles throughout the simulations. As a sanity check, we overlay the analytical solution of $4/3$-polytrope (purple dahsed line).
    }
    \label{fig:resoln_star}
\end{figure}

A typical choice for resolutions of hydro-simulations of TDEs or micro-TDEs is $N\sim10^5$ particles \citep[e.g.][]{Mainetti2017,Kremer2022}.
We performed resolution tests to determine whether or not a higher resolution is needed, by using \texttt{PHANTOM} to model the initial stellar profile using different numbers of SPH particles $N=10^5, 2\times10^5, 4\times10^5, 8\times10^5, 10^6$. 
In particular, we compare the radial density profiles of the fully relaxed $1M_{\odot}$ star with the numbers of SPH particles given above in Fig.~\ref{fig:resoln_star}. The gray region shows where the initial profile varies the most, which occurs at the surface of the star. We find that different resolutions only cause the density to fluctuate by $\sim$0.01\%, which only takes place in less than 1\% of the SPH particles by mass and $\lesssim0.2R_{\odot}$ by radius. Overall, the density profiles for resolutions from $N=10^{5}$ to $N=10^{6}$ particles show excellent agreement. Therefore, we run all TPE simulations, starting from their stellar profiles, with particle number $N=10^5$. 
As a comparison, we also depict the polytropic star with $\gamma=4/3$ of the same mass using a purple dashed line. 
% 
% Finally, we test that the key quantities in the analysis of our TPE simulations, described in \S~\ref{sec:analysis}, show reasonable degree of convergence. \noteTH{So in TPE simulations? the key quantities in the analysis section are relevant for TPEs.}\cx{yes}

\section{Morphology of TPE} \label{sec:morphology}
% \noteTH{It seems the main characteristic feature of TPEs due to relatively gradual mass stripping with angular momentum comparable to that of the stellar orbit are three, 1) the formation of spirals, 2) possible debris-star interactions, and 3) efficient circularization of debris into an accretion disk. All these features are found in Figure \ref{fig:morphology1}. In the first paragraph of the current version of this section, it says there are unique features but does not explain exactly what they are. Instead, how about explaining what the characteristic features are first (1st paragraph) and use Figure \ref{fig:morphology1} to illustrate the features (paragraphs 2 and 3)?   }

\begin{figure*}
    \centering
    \includegraphics[width=0.66\columnwidth]{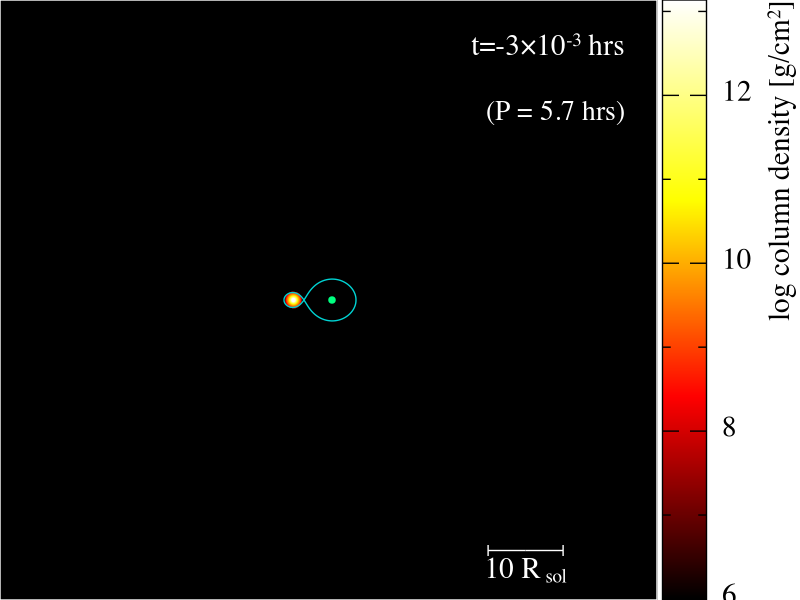}
    \includegraphics[width=0.66\columnwidth]{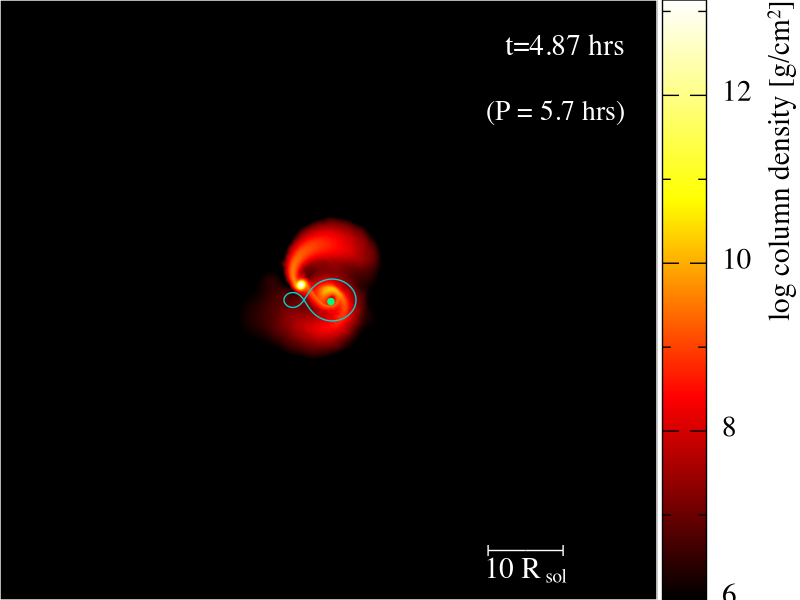}
    \includegraphics[width=0.66\columnwidth]{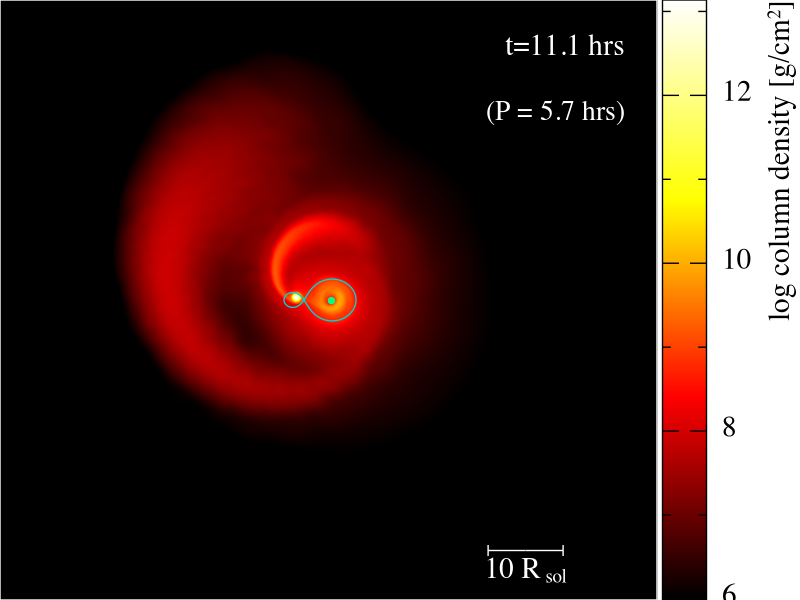} \\
    \includegraphics[width=0.66\columnwidth]{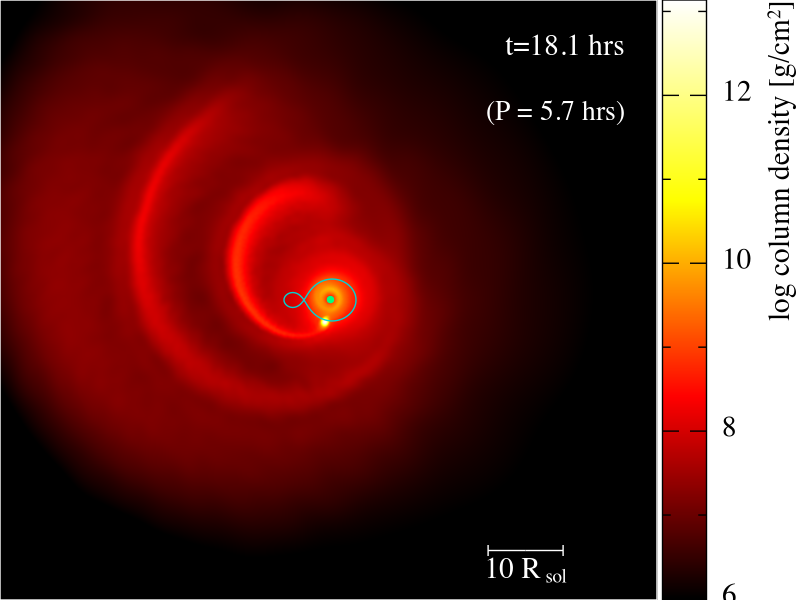}
    \includegraphics[width=0.66\columnwidth]{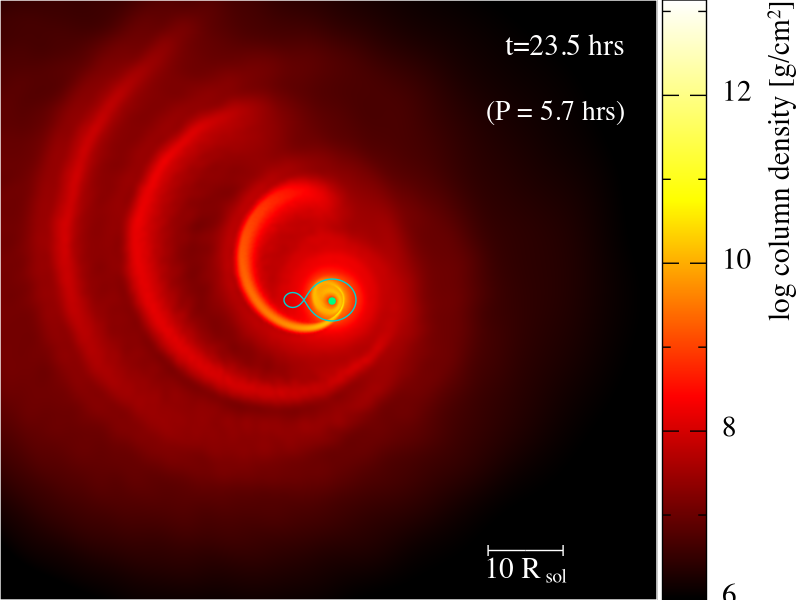}
    \includegraphics[width=0.66\columnwidth]{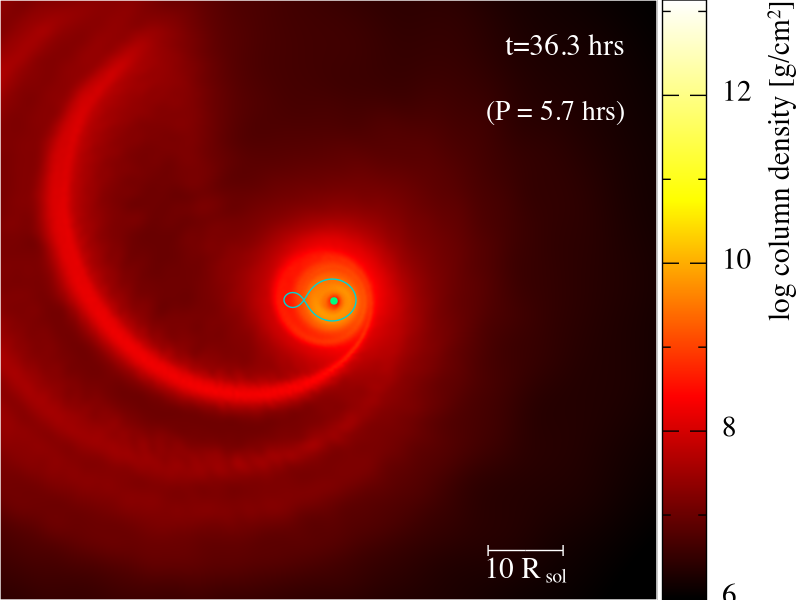}
    \caption{$\mathcal{M}(1,0.4,1)$ -- Tidal peeling morphology of a $1M_{\odot}$ star and a $10M_{\odot}$ BH, where the orbit is  initially a low-eccentricity inspiral ($e_0=0.4$), and the pericenter distance between star and BH is 1 tidal radius ($r_p=2.2R_{\odot}$; $\beta=1$). The color bar shows the projection of log-scale column density in the x-y plane. We overlay the {\it initial} equipotential surface of the binary to show that the stellar material fills up the Roche Lobe around the BH, and the star loses mass through the Lagrangian points. The initial orbital period is quoted in parathesis, specifically, $P\approx5.7$ hours in this model. We show six time frames of the event that demonstrate the tidal ``peeling" process, until the star is completely disrupted by the BH. The star orbits around the BH and passes through the pericenter four times until it is torn apart by the BH. 
    }
    \label{fig:morphology1}
\end{figure*}

\begin{figure*}
    \centering
    \includegraphics[width=0.66\columnwidth]{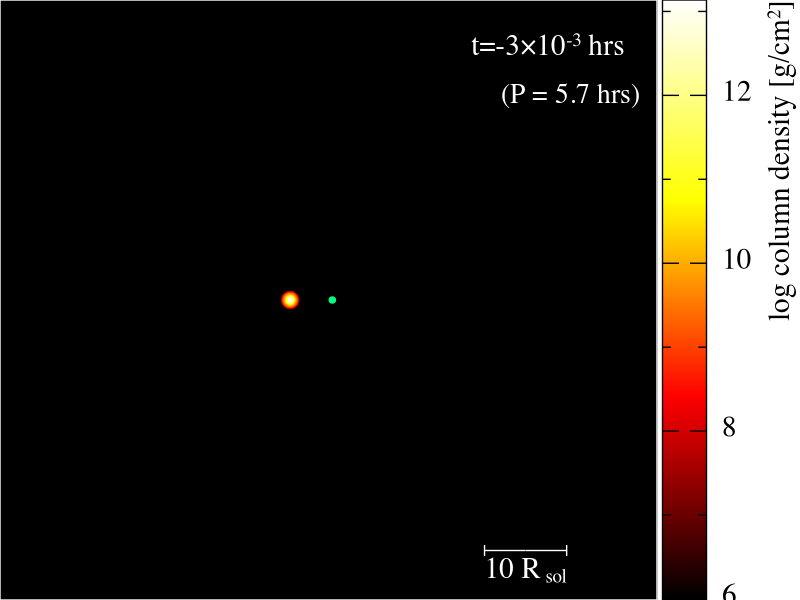}
    \includegraphics[width=0.66\columnwidth]{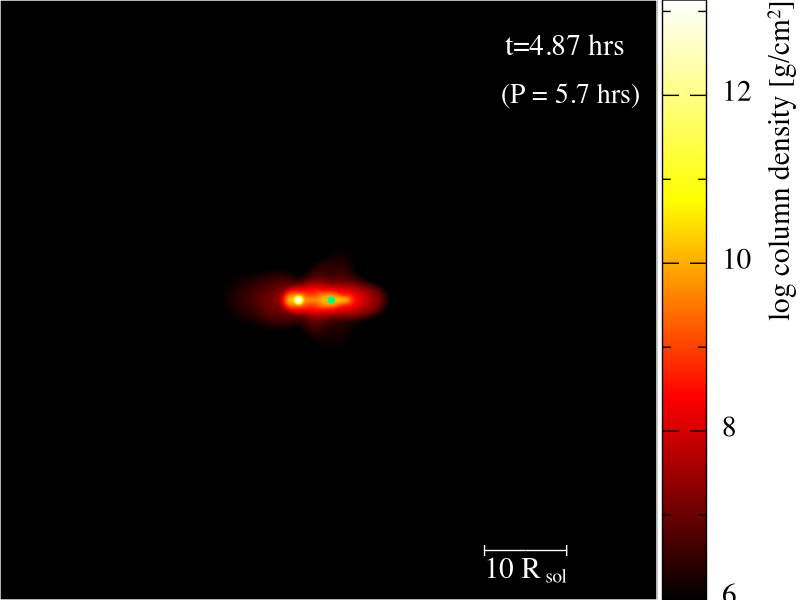}
    \includegraphics[width=0.66\columnwidth]{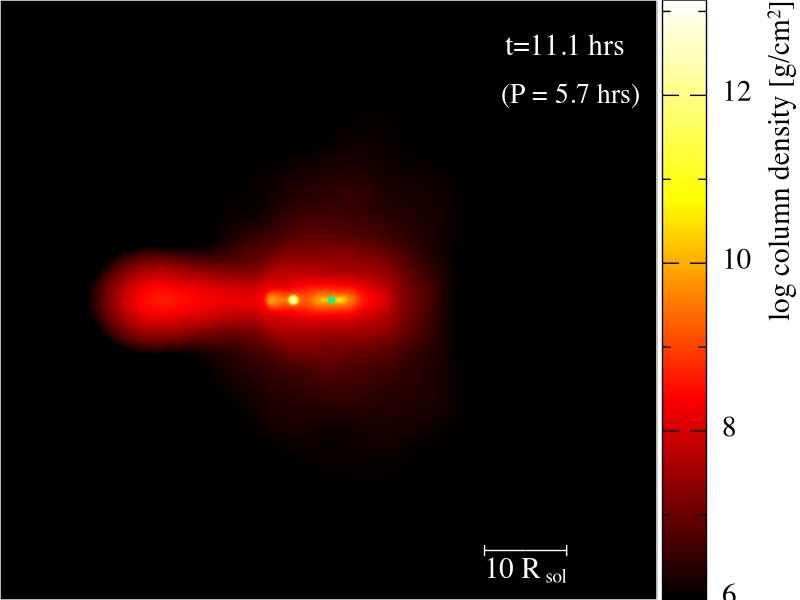} \\
    \includegraphics[width=0.66\columnwidth]{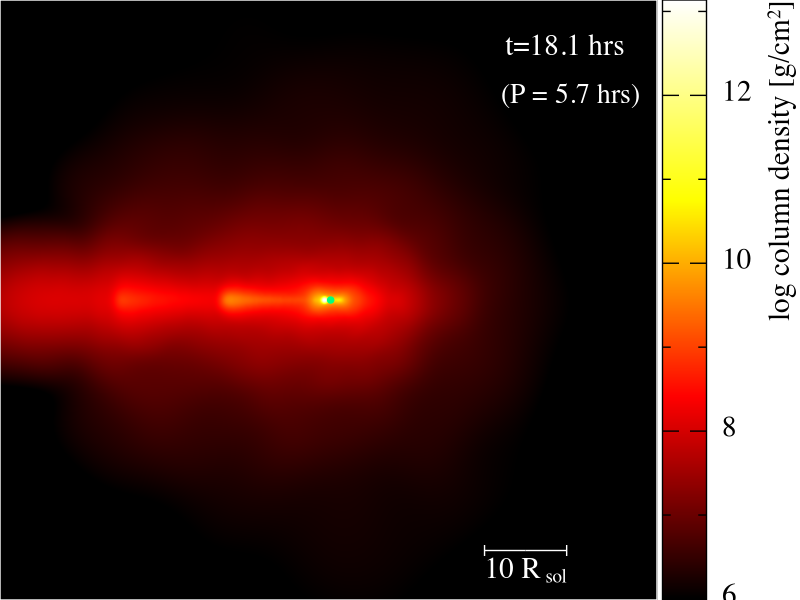}
    \includegraphics[width=0.66\columnwidth]{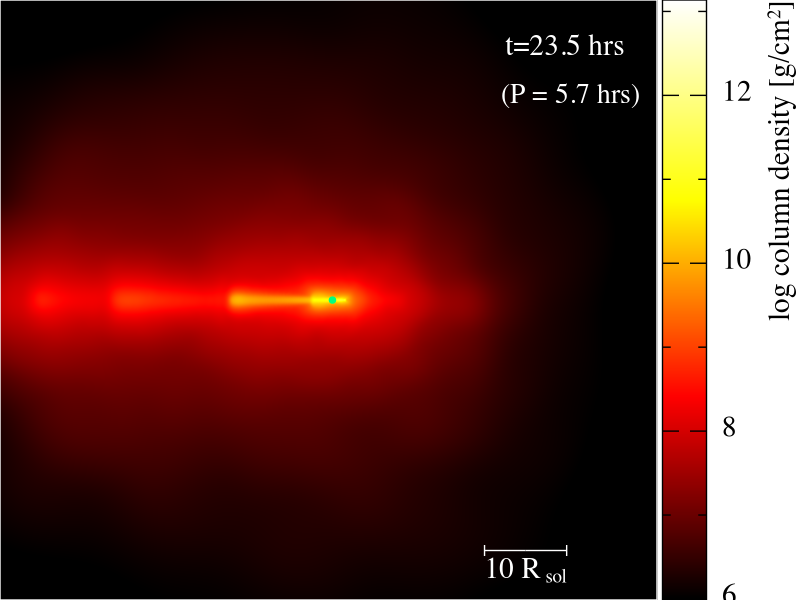}
    \includegraphics[width=0.66\columnwidth]{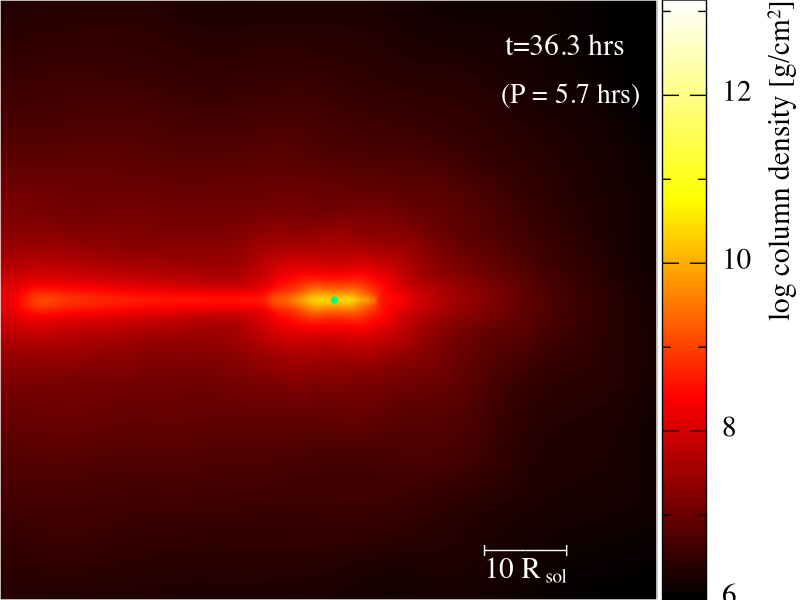}
    \caption{$\mathcal{M}(1,0.4,1)$ -- Same snapshots of the simulation as in Fig.~\ref{fig:morphology1} on the x-z plane, or edge-on view of the orbit and the accretion disk. }
    \label{fig:morph-xz}
\end{figure*}
The stars in our TPE simulations encounter the BH in low-eccentricity ($e=0-0.6$) and ultra-compact ($\beta=0.25-1$) orbits. Depending on the initial conditions, the mass of the star can be slowly peeled by the BH, and stellar material is lost on the timescale of many orbital periods. In general, TPEs will have novel morphological evolution, e.g. distinct morphology from that seen in TDEs or micro-TDEs, and in particular, 1) gradual tidal stripping and formation of spirals, 2) possible debris-star interactions, and 3) efficient circularization of debris into an accretion disk. Each of these is demonstrated in the following examples.

Fig.~\ref{fig:morphology1} shows a typical morphology of a TPE, where the column density of the gas particles is shown in the color bar and the BH is represented by the green dot. 
In this example (Model $\mathcal{M}(1,0.4,1)$; recall the definition in \S~\ref{sec:analysis}), the $1M_{\odot}$ star on an eccentric orbit with $\beta=1$ is ``peeled'' due to the tidal influence of the BH, which continues for four orbits before the star is totally disrupted (at the $\sim 4^{\rm th}$ orbits). The snapshots are taken at $t=0, 4.9, 12.0, 18.2, 23.5$ and $36.3$ hours since the onset of the simulation, where the orbital period is $P\approx5.7$ hrs. Some stellar debris circularizes and forms an accretion disk around the BH, while some becomes unbound and are ejected into infinity, including mass lost through the ``L3" point; we show the initial equipotential surface of the binary in each panel.
% \cx{overlay equipotential surface and elliptical orbit}. 
This can be more clearly seen in Fig.~\ref{fig:morph-xz} that shows the edge-on view of Fig.~\ref{fig:morphology1}. The disk is initially smaller than the pericenter distance of the orbit for a short period of time, before it inflates and puffs up later on due to radiation pressure and shock heating, similar to the findings of \citet{Wang2021}.  

\begin{figure*}
    \centering
    \includegraphics[width=0.66\columnwidth]{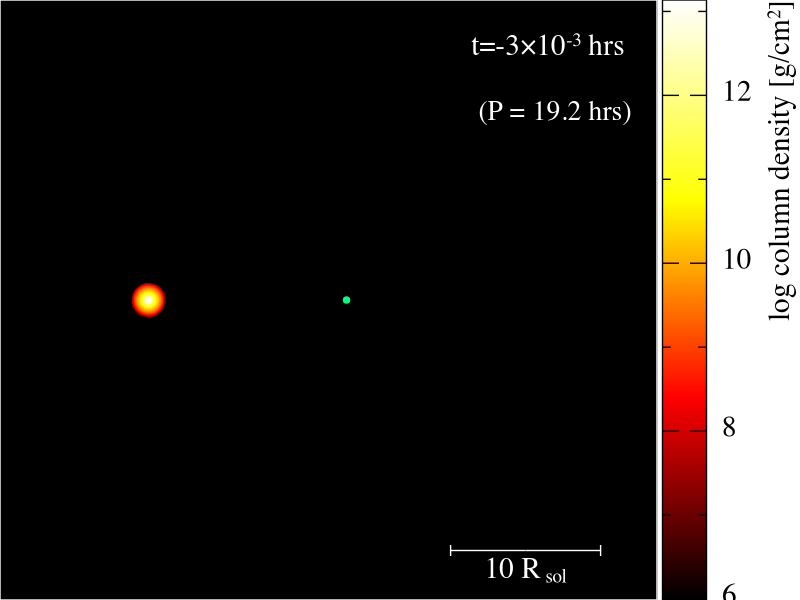}
    \includegraphics[width=0.66\columnwidth]{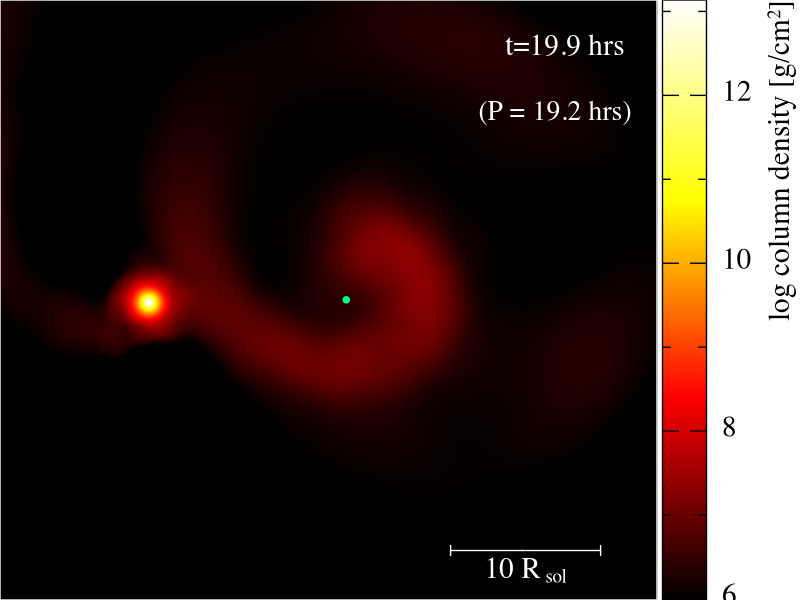}
    \includegraphics[width=0.66\columnwidth]{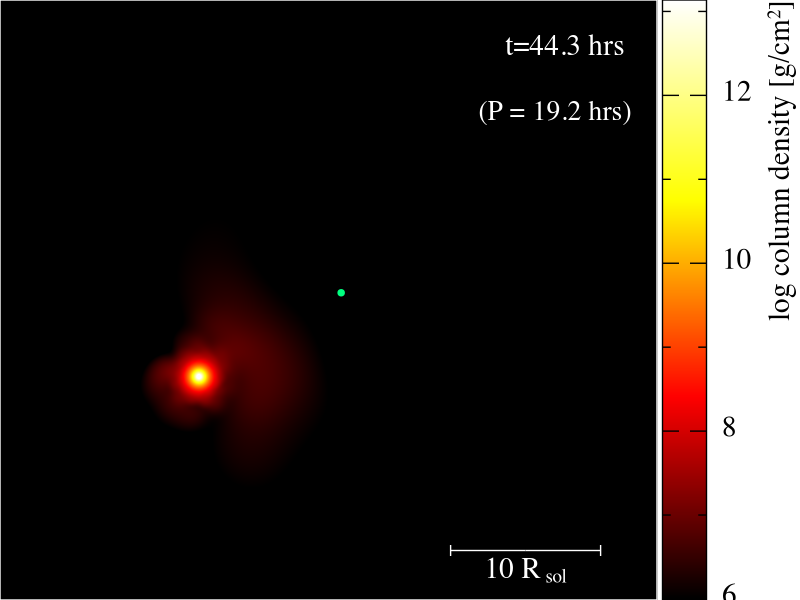} \\
    \caption{$\mathcal{M}(1,0.6,0.67)$ -- Tidal peeling of the same BH-star binary described in Fig.~\ref{fig:morphology1}, but with initial eccentricity $e_0=0.6$ and $\beta=0.67$. The initial orbital period (in parenthesis) is 19.2 hours.}
    \label{fig:morph-shock}
\end{figure*}

Generally, tidal peeling is more violent for smaller orbital separations. 
% The EM emission of a TPE should be modulated by the accretion rate onto the BH. 
All of our TPE simulations result in super-Eddington BH accretion rates. However, 
a significant fraction of the star being tidally disrupted, leaving most of the dense stellar material around the BH, results in large optical depth that likely will delay and dim the EM emission from a TPE. In reality, the luminosity could be modulated by several mechanisms such as jet emission or wind outflow from the accretion disk, which are not included in this study.
% Unfortunately these phenomena cannot be resolved in our simulations. We will discuss the implication of the TPEs accretion rates in the upcoming sections. 
Additionally, in some configurations, such as $\mathcal{M}(1,0.6,0.67)$ in Fig.~\ref{fig:morph-shock}, the star intersects with its own tidal streams periodically, which will form a shock front that further modifies the luminosity from the TPE. In this model, the remnant remains intact for many orbits.
In the second panel, the star encounters the tail of its own stream formed in the last orbit, leaving behind a hot ploom near the star as seen in the last panel. 
Although these phenomena cannot be resolved in our simulations, in the following sections, we will qualitatively discuss their implications for the overall EM signature of the TPEs in addition to the accretion rates that we measure directly from the simulations.

\begin{figure*}
    \centering
    \includegraphics[width=0.66\columnwidth]{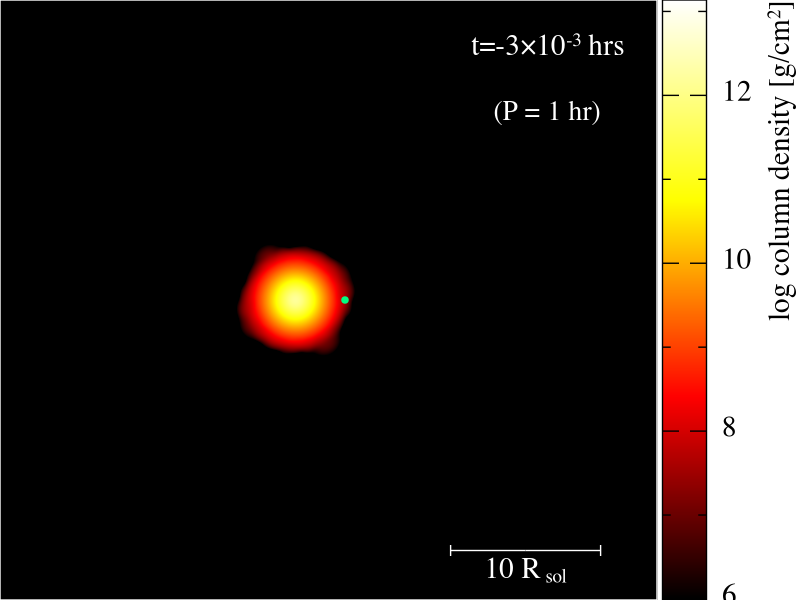}
    \includegraphics[width=0.66\columnwidth]{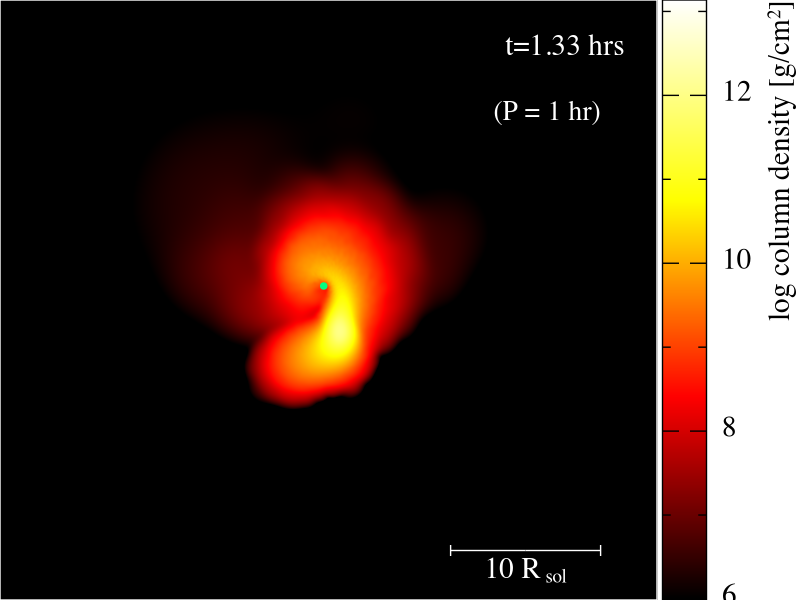}
    \includegraphics[width=0.66\columnwidth]{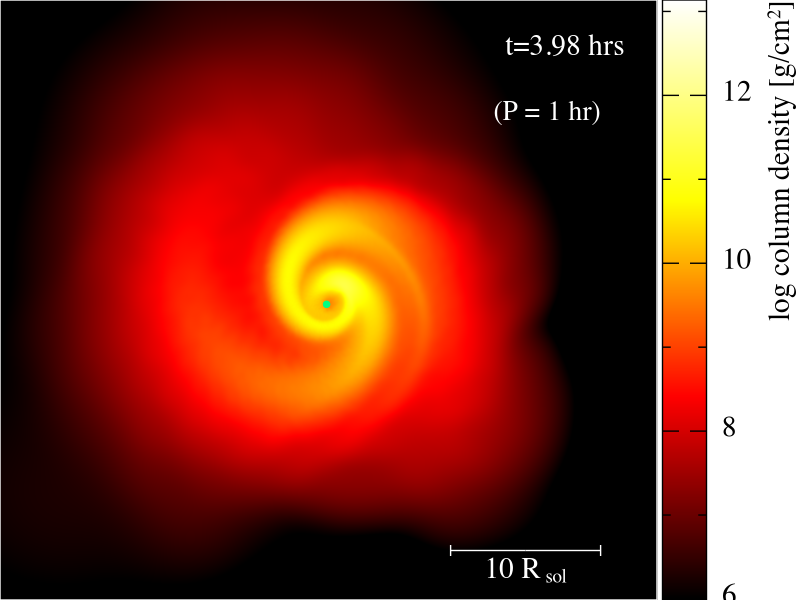} \\
    \includegraphics[width=0.66\columnwidth]{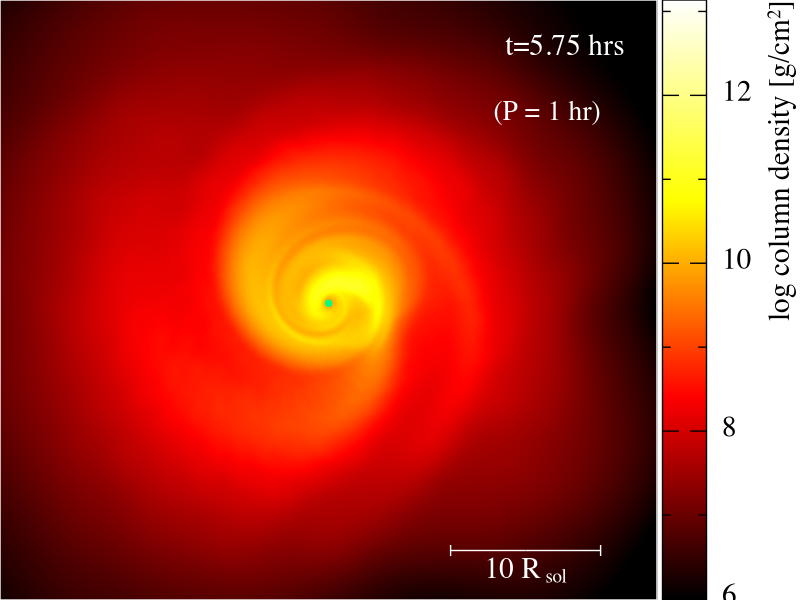}
    \includegraphics[width=0.66\columnwidth]{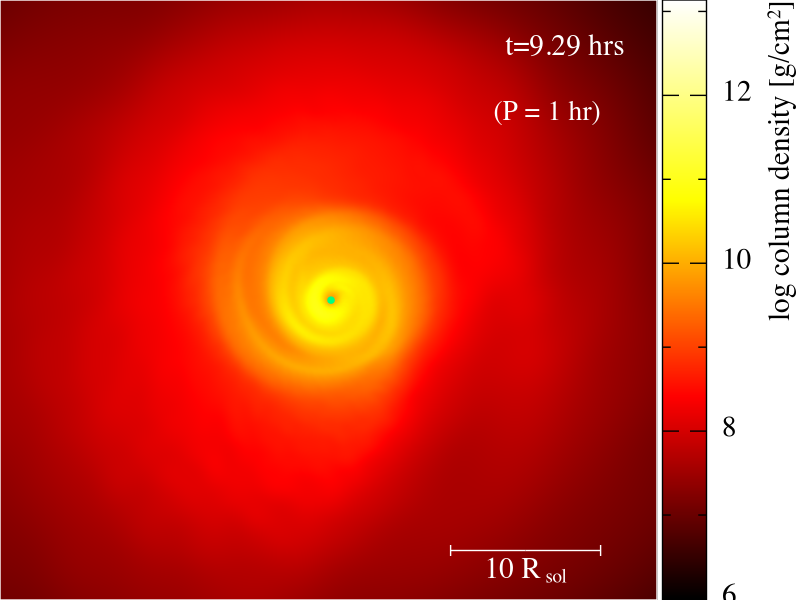}
    \includegraphics[width=0.66\columnwidth]{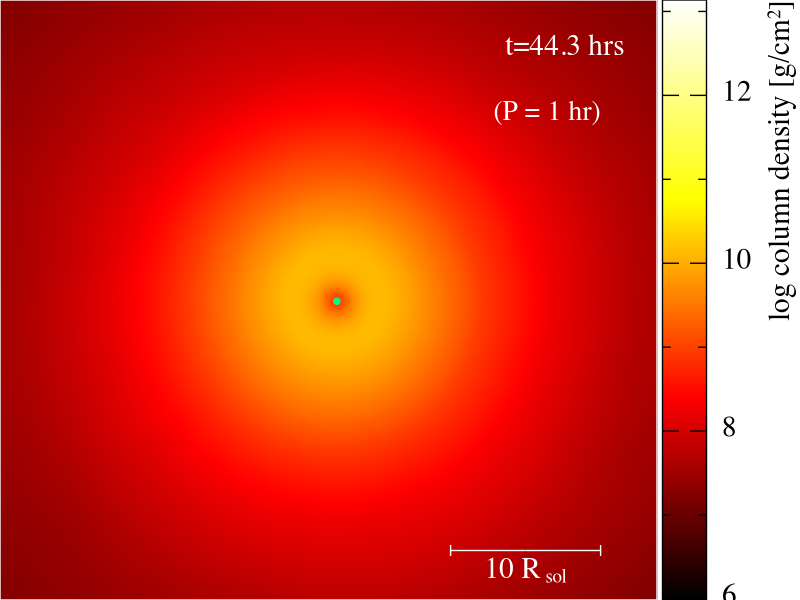}
    \caption{$\mathcal{M}(5,0.0,1)$ -- Tidal peeling of BH-star with a higher stellar mass, $M_{\rm s}=5M_{\odot}$, initially circular orbit ($e_0$=0) and pericenter distance equal to the tidal radius $\beta=1$. The initial orbital period of the binary is $\sim1$ hour. The star is completely disrupted soon after the beginning of the simulation. }
    \label{fig:morphology4}
\end{figure*}

% \noteTH{It seems this paragraph belongs to Section~\ref{sec:massive_stars}}
Finally, TPEs from the interaction of BHs with more massive stars are considered since
stars near the galactic center \citep{Genzel2003,Levin2003,Paumard2006} and those formed in an AGN disk \citep{Levin2003,Goodman2004} are also thought to be preferentially massive, and they offer morphology different from TPEs with a solar-like star. In Fig.~\ref{fig:morphology4}, we demonstrate the TPE between a $5M_{\odot}$ star and the BH in circular orbit with the initial separation of one tidal radius. The surface of this star is almost in contact with the BH, $a=r_p\approx 1.3 R_s$. Compared to a solar-like star in the same initial orbit, a more massive star experiences more rapid tidal peeling. As a result, the spirals formed from the disrupted material are more closely packed, compared to those in Fig.~\ref{fig:morphology1}. The snapshots of the TPE are taken at $t=0, 0.88, 1.77, 2.66, 3.54$ and $4.43$ hours, and this TPE model has orbital time $P\approx 1$ hr. The massive star is totally disrupted within the first orbit, and the stellar material eventually circularizes into a smooth disk. 

\section{Accretion rate and orbital evolution of TPEs} \label{sec:em_features}
% \addTH{This section is substantially long compared to other sections. How about we split this into two subsections? For example, section 6.1. Overview (paragraphs 1,2,3), section 6.2 Dependence on the initial parameters? I added those section titles. Feel free to remove or edit.}

\begin{figure*}
    \centering
    \includegraphics[width=\columnwidth]{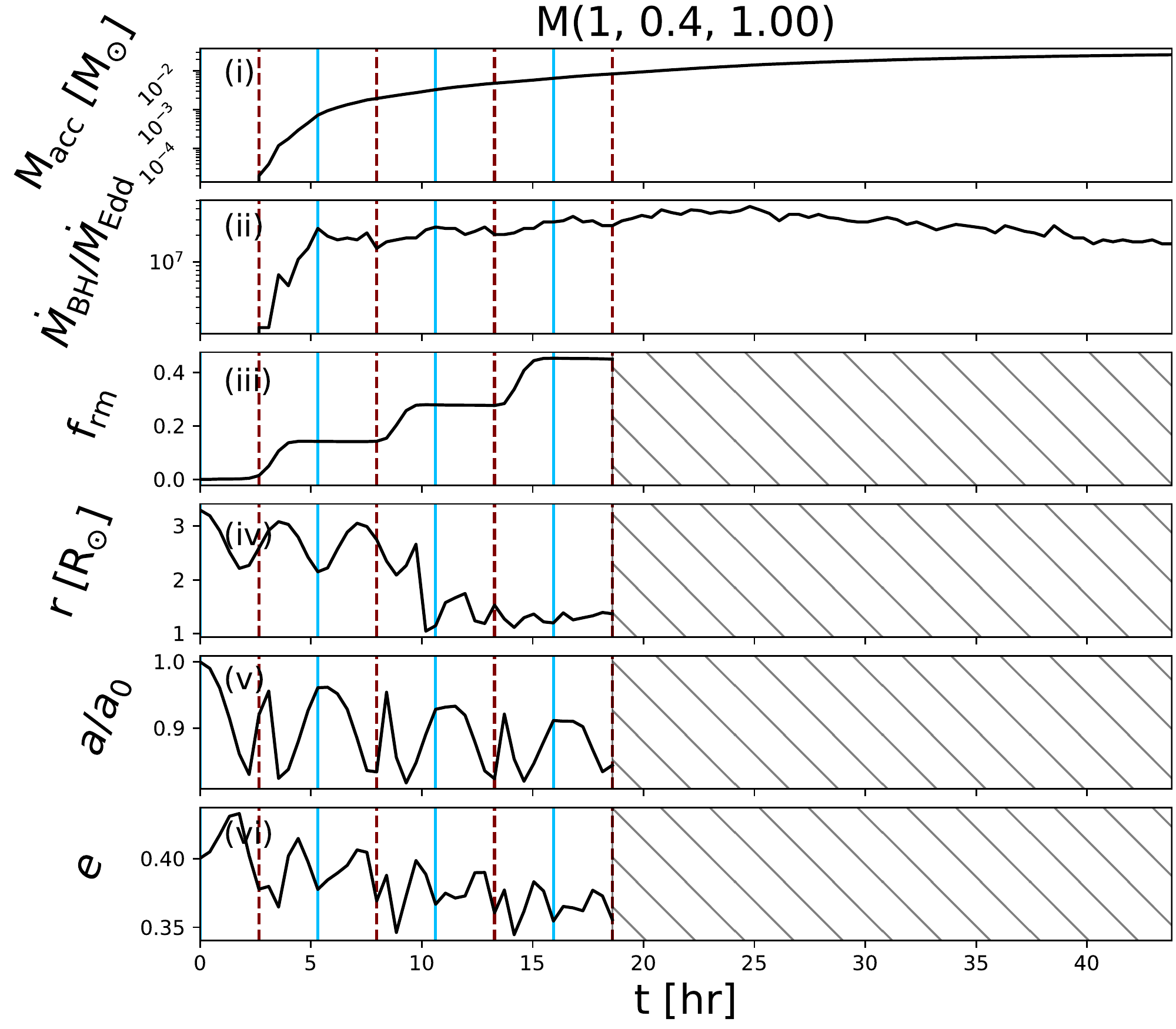} 
    \includegraphics[width=\columnwidth]{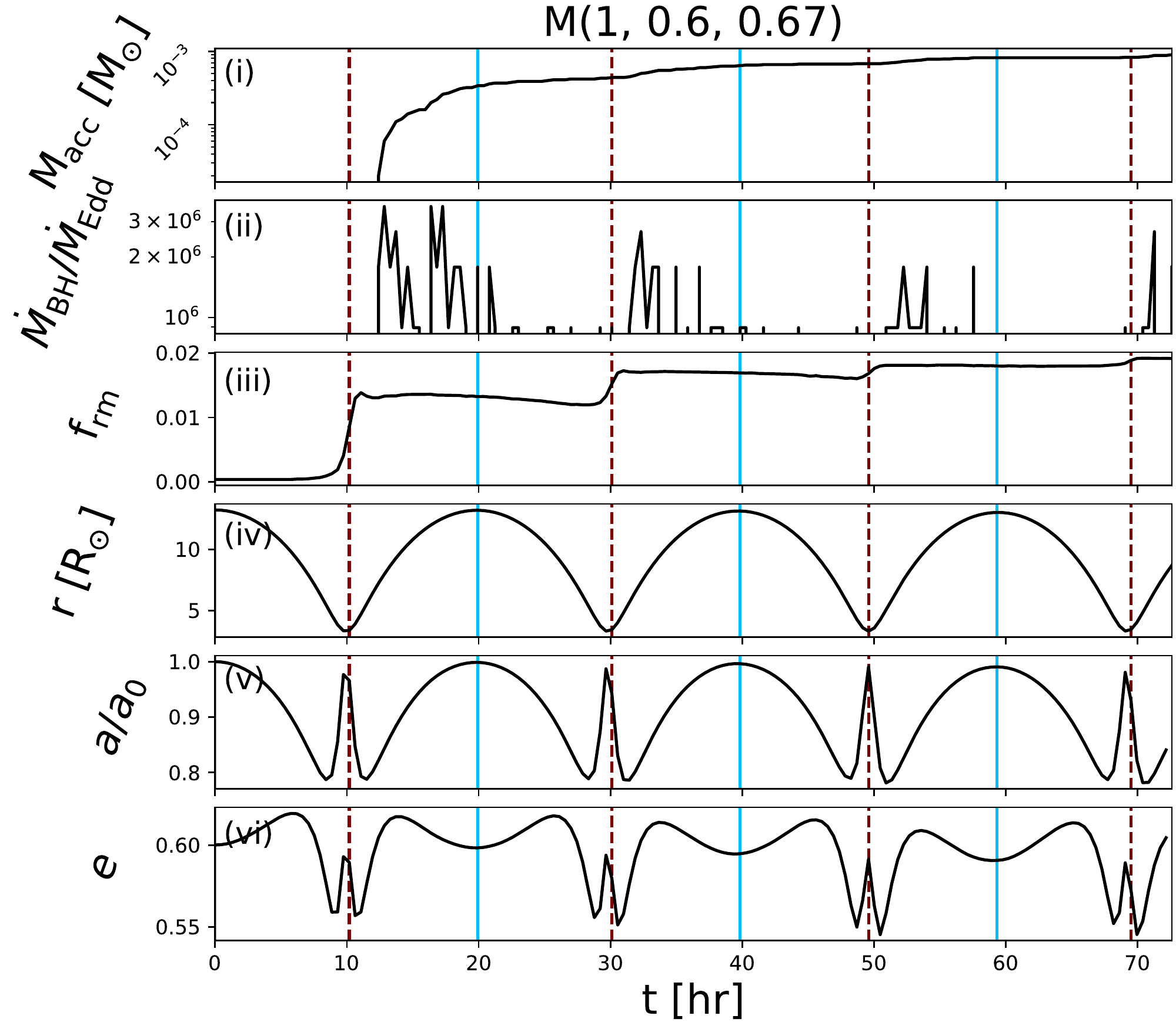}
    \caption{Time evolution of key physical quantities characterizing TPEs, for the models $\mathcal{M}(1,0.4,1)$ (left) and $\mathcal{M}(1,0.6,0.67)$ (right). The six panels, from top to bottom, show the (i) mass accreted by the BH, (ii) accretion rate in Eddington units, (iii) the fraction of mass removed from the star, (iv) the separation between the remnant and the BH, (v) the evolution of the semi-major axis and (vi) the eccentricity. 
    The pericenter and apocenter passages are labeled with red-dashed and blue-solid lines, respectively. The hatched regions represent total disruption of the star. 
    }
    \label{fig:6-panel1}
\end{figure*}

\subsection{Overview using two examples}

% Our simulations offer estimates for the accretion rate and orbital evolution of TPEs.
% 
Fig.~\ref{fig:6-panel1} demonstrates six key features of mildly eccentric TPEs for the case of the 10$M_{\odot}$ BH and the $1M_{\odot}$ star. This figure presents two models -- $\mathcal{M}(1,0.4,1)$ (left): initial eccentricity ($e_0$) is 0.4 and initial pericenter distance $r_p/R_t$=1 ($\beta=1$), and $\mathcal{M}(1,0.6,0.67)$ (right): a more eccentric and less compact model with $e_0$=0.6 and $r_p/R_t$=1.5 ($\beta=0.67$). We show the time evolution of (i) the mass accreted onto the BH ($M_{\rm acc}$), (ii) the mass accretion rate ($\dot{M}_{\rm BH}$) in Eddington luminosity $\dot{M}_{\rm Edd}=L_{\rm Edd}/0.1c^2$, (iii) the fraction of mass removed from the star ($f_{\rm rm}$), (iv) the orbital separation ($r$), (v) the evolution of the SMA normalized to its initial value ($a/a_0$) and (vi) the evolution of eccentricity  ($e$). The bottom four panels of Fig.~\ref{fig:6-panel1} reflect the properties of the stellar remnant and are therefore only computed before total disruption; the time after total disruption of the star is labeled with hatched lines. Finally, we show the times of pericenter and apocenter passages with red-dashed lines and blue-solid lines, respectively. 

In the first model, $\mathcal{M}(1,0.4,1)$, the mass of the BH grows monotonically with time, while the accretion rate increases until a plateau around 5 hours ($\sim P$), exceeding the Eddington limit by more than seven orders of magnitude. 
In fact, the values of $\dot{M}_{\rm BH}$ that we find are typically super-Eddington within the first few orbits of disruption, if $r_p$ within $\sim3 R_t$.
In this model, the stellar remnant orbits around the BH on a $\sim5$hr orbital timescale, during which the binary separation shrinks and the fraction of stellar mass removed becomes larger until the star gets totally disrupted after approximately 4 orbital times.  The large fluctuations in $a$ and $e$ indicate that the star-BH orbit is not Keplerian due to tidal effects and shocks, resulting in the dissipation of orbital energy and asymmetric mass loss. 

For an initially less compact binary, e.g. $\mathcal{M}(1,0.6,0.67)$ (right-hand side of Fig.~\ref{fig:6-panel1}), the stellar remnant does not undergo total disruption in the first few orbits. In fact, the mass accretion rate spikes after each pericenter passage (minima in $r$) with a small time delay, while the peak level decreases over time.  Similar observations have been reported in simulations of binary stars, where the peak of mass transfer rate is found shortly after each binary orbit's pericenter \citep{Lajoie2011}. $a$ and $e$ show fluctuations unique to TPEs, discussed further in \S~\ref{sec:discussion}, indicating non-Keplerian orbital evolution, even for a slightly tidally disrupted star's orbit.

\subsection{Dependence on the initial conditions}

In this section, we investigate the dependence of the six key quantities above on different initial conditions, namely $M_s$,  $e_0$, and $\beta$, providing characteristics of the EM emission of TPEs.  We measure these quantities during the first three orbits of the remnant around the BH, from one apocenter to the next (between blue solid lines in Fig.~\ref{fig:6-panel1}). In particular, we compute the change per-orbit of mass accreted onto the BH, the BH accretion rate, and the fractional stellar mass removed, which are denoted by $M_{\rm acc,a}$, $\dot{M}_{\rm BH,a}$ and $f_{\rm rm,a}$, respectively. This allows us to take into account any enhancements
 in $\dot{M}_{\rm BH}$ during each orbit, including the peaks near the pericenters as seen in the right-hand side of Fig.~\ref{fig:6-panel1}. We also evaluate the total change in SMA ($\Delta a/a_0$) and eccentricity ($\Delta e$) each orbit.

% \subsection*{Black hole mass accretion and stellar mass loss} \label{subsec:macc_mdot}

\begin{figure*}
    \centering
    \includegraphics[width=2\columnwidth]{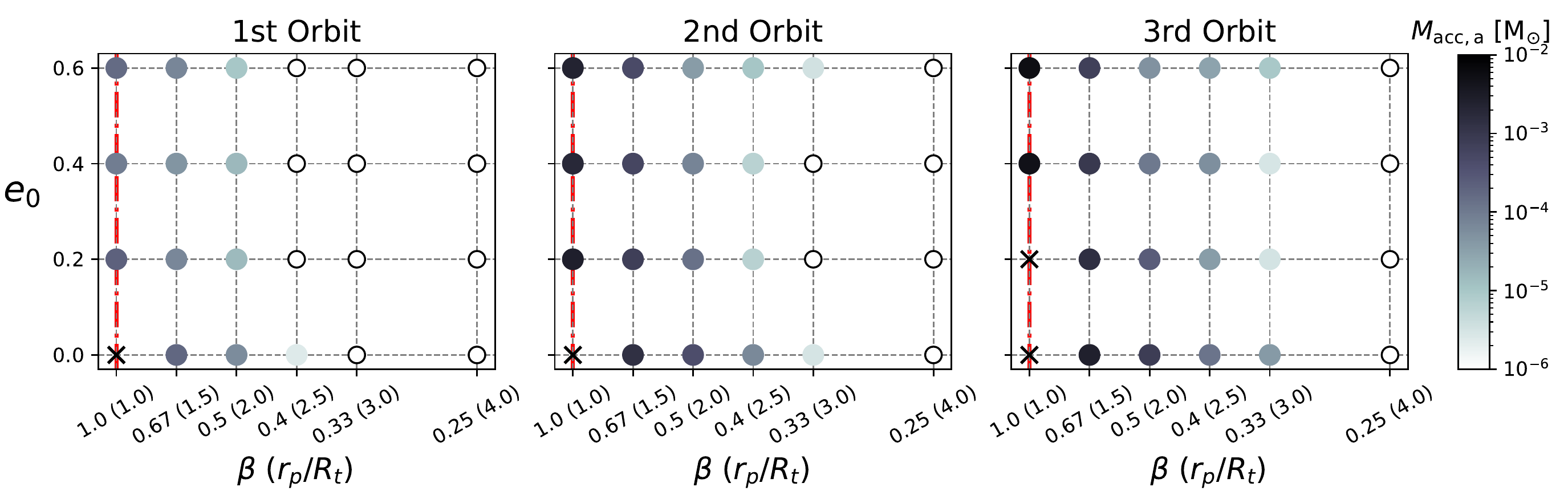}
    \caption{The change of mass accretion onto the BH per orbit, $M_{\rm acc,a}$, for TPEs with a $1M_{\odot}$ star, as a function of initial $e_0$ and $\beta$, evaluated for the first three orbits of the stellar remnant around the BH. We show the pericenter distances corresponding 
    to each $\beta$ in the parentheses. The darker end of the color bar represents larger $M_{\rm acc,a}$ values, which decrease as the initial orbit becomes wider and more eccentric. In the most compact configurations, the star is totally disrupted (crosses), while in the least compact orbits, zero mass is accreted by the BH (open circles). The onset of mass transfer is analytically expected to occur when $r_p\approx R_t$ (red dotted line).}
    \label{fig:macc}
\end{figure*}

In comparison with a typical TDE or micro-TDE, where the star is on a parabolic orbit and more than half of its mass can be lost at the first pericenter passage \citep[e.g.][]{Mainetti2017,Bartos2017,Yang2020,Kremer2022}, in a TPE, the star typically loses mass to the BH more gradually over many orbits around the BH. The degree of mass loss from the star and the mass accretion onto the BH can be different, depending on the choices of $M_s$, $e_0$, and $\beta$. 

Fig.~\ref{fig:macc} shows the orbital change of mass accretion, $M_{\rm acc,a}$, of TPEs with the $1M_{\odot}$ star and the $10M_{\odot}$ BH, under different assumptions for $\beta$ (x-axis) and $e_0$ (y-axis).  In the most compact models ($\beta=1$), the star gets totally disrupted within the first three orbits, which are denoted with crosses. This is roughly consistent with the analytical expectation that the star undergoes tidal disruption when the pericenter distance of the orbit is comparable to the tidal radius, i.e. $r_p/R_t\sim1$ (red dot-dashed line). More generally, $M_{\rm acc,a}$ is larger for initially more compact orbits, meaning smaller $r_p$ (larger $\beta$) and smaller $e_0$. The latter is equivalent to having smaller initial orbital separation, since we initially place the star at the apocenter distance $r_{\rm apo}=a_0(1+e_0)$. However, we see a smaller dependence of $M_{\rm acc,a}$ on the initial eccentricity than on initial pericenter distance. The amount of mass accreted by the BH inevitably increases over time once mass transfer begins, resulting in the highest values of $M_{\rm acc,a}$ in the third orbit.  In the models with the largest pericenter distances, $r_p/R_t\gtrsim3$, there is no mass accretion onto the BH in the first three orbits, denoted by the open circles. 

\begin{figure*}
    \centering
    \includegraphics[width=2\columnwidth]{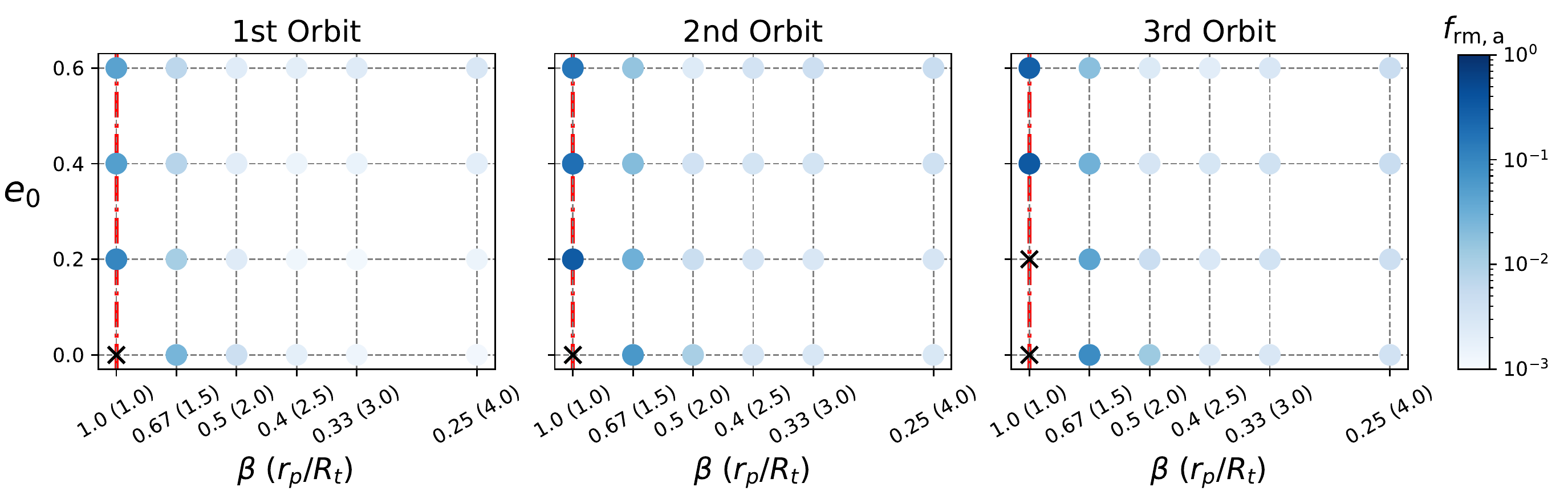}
    \caption{Similar to Fig.~\ref{fig:macc}, but we show the orbit-averaged fraction of mass removed from the star and the BH, $f_{\rm rm,a}$ = $M_{\rm rm}/M_s$, where $M_{\rm rm}$ = $M_s - M_{\rm rem, s} + M_{\rm acc,BH}$, where $M_s$ is the total mass of the star, $M_{\rm rem, s}$ the remnant mass and $M_{\rm acc,BH}$ the mass accreted onto the BH. The crosses represent total disruption. The red dot-dashed lines again represent the onset of mass transfer limit at $r_p\approx R_t$.}
    \label{fig:m_rm}
\end{figure*}

We see similar trends in the fraction of stellar material removed from the star (or $f_{\rm rm,a}$; Fig.~\ref{fig:m_rm}). Tidal peeling can remove stellar mass slowly over a few orbital times, which can be seen from the persistent increase of $f_{\rm rm,a}$ over the first three orbits. Generally, a larger fraction of the star is removed when the initial orbit has smaller $r_p$ and $e_0$, and as time goes on. Note that even in the widest binaries ($r_p/R_t\gtrsim3$), a small amount of stellar mass is removed under tidal effects, which is beyond the analytical prediction \citep[e.g.][]{Zalamea2010} for the onset of mass loss (red dot-dashed line), although the mass accretion onto the BH can be zero (as seen in Fig.~\ref{fig:macc}). Finally, we again observe larger variations in $f_{\rm rm,a}$ due to $r_p$ than due to $e_0$.

\begin{figure*}
    \centering
    \includegraphics[width=2\columnwidth]{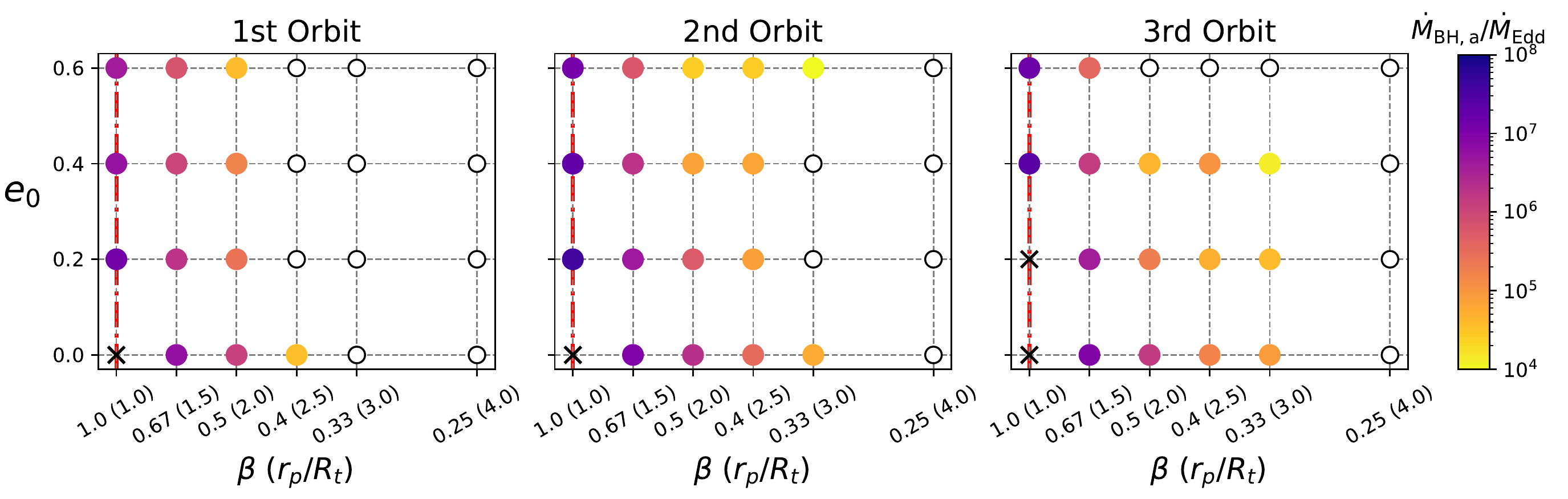}
    \caption{Mass accretion rate onto the BH, $\dot{M}_{\rm BH}$ as a function of initial eccentricity and penetration factor (the corresponding pericenter distances are quoted in parentheses), taken at the 1st to 3rd pericenter passages. $\dot{M}_{\rm BH}$ are given in units of $\dot{M}_{\rm Edd}\sim2.2\times10^{-7}M_{\odot}$/yr for a $10M_{\odot}$ BH. The open circles represent zero accretion rate. The open circles represent mass accretion rate of zero. The crosses represent total disruption. }
    \label{fig:mdot}
\end{figure*}

% The observed luminosity of TPEs can be modulated by the mass accretion rate onto the BH. 
Fig.~\ref{fig:mdot} shows that typically $\dot{M}_{\rm BH,a}$ range from $\sim10^4$ to $10^8$ times the Eddington accretion rate of the BH. The values of $\dot{M}_{\rm BH,a}$ are overall higher when the initial binary orbit is more compact and less eccentric, although, like in Fig.~\ref{fig:macc} and \ref{fig:m_rm}, the impact of the initial value of $r_p$ is larger than the impact of $e_0$. Like the trend in both BH mass accretion and fraction of stellar mass loss, the values of $\dot{M}_{\rm BH,a}$ tend to increase over time, except in some models with $e_0$=0.6, e.g. $\mathcal{M}(1,0.6,0.67)$ in Fig.~\ref{fig:6-panel1}, where the tidal influence is the weakest due to the large initial separation between the star and the BH.

Most TPE models in our simulations indicate partial disruption of the star, which suggests EM emission from TPEs persisting over many orbital times. Although we only simulate the first few orbital times of TPEs in this work, we investigate the orbital evolution of the stellar remnant during this time, and we attempt to find patterns in the evolutions of the SMA and the eccentricity that could predict whether the binary separation widens or becomes more compact. Future work should investigate the long-term behavior of star-BH TPEs, in order to determine (1) the full duration of their EM emission, and (2) whether or not the star will be eventually totally disrupted by the BH. 
% The latter will affect the number of stars and compact objects predicted through the AGN formation channel.

\begin{figure*}
    \centering
    \includegraphics[width=2\columnwidth]{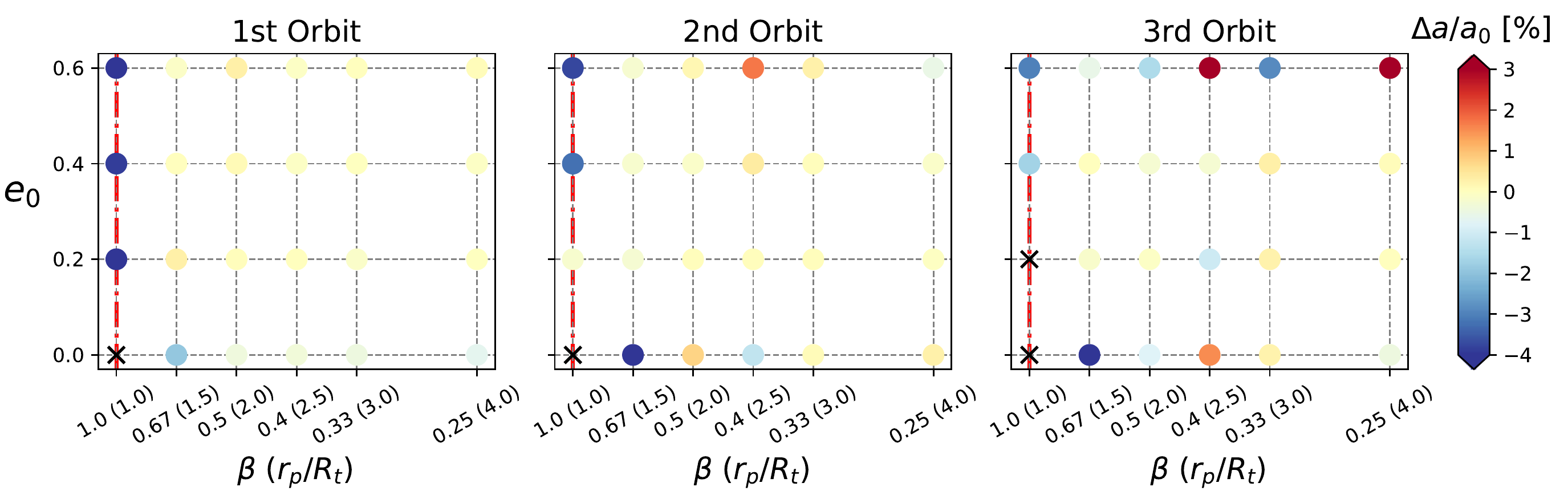}
    \caption{The change of semi-major axis ($\Delta a$) normalized by its initial value ($a_0$) during the first, second and third orbit around the BH, given in percentages. The yellow colors represent (near) zero changes in SMA during the orbit, while the redder (bluer) points represent orbit expanding (shrinking). }
    \label{fig:del_a}
\end{figure*}

In Fig.~\ref{fig:del_a}, we demonstrate the variations in SMA ($\Delta a$) per orbit evaluated during the first three orbits of TPEs with the $1M_{\odot}$ star around the BH. We investigate the change in $\Delta a$, normalized by the initial SMA $a_0$ of each model, due to different initial conditions $\beta$ and $e_0$. The color bars show percentage values of $\Delta a/a_0$, which typically fluctuate within $\sim$4\%. We observe that in most models, $\Delta a$ remains roughly zero (yellow points), corresponding to very small variation in $a$ during one orbit, meaning that the orbital separation at one apocenter is not too different from the next one. The redder points in Fig.~\ref{fig:del_a} correspond to the models where the orbits are widening ($\Delta a > 0$); the bluer points corresponds to shrinking orbits ($\Delta a < 0$). There is a lack of overall trends that dictates whether $\Delta a$ increases or decreases with the two initial conditions, except that the most compact orbits tend to decay. 

% In the last orbits, however, the models with the highest $e_0$ (e.g. top left two models) are circularizing (seen also in Fig.~\ref{fig:del_e}), and their orbital separations are reducing.
% Similarly, the models with initially circular orbits (bottom points in the third panel) are widening, when initial pericenter distance is small (within $\sim$2.5 tidal radii), which can be due to the fact that these models lose the most stellar mass in each orbit, see the corresponding points in Fig.~\ref{fig:m_rm}. 

\begin{figure*}
    \centering
    \includegraphics[width=2\columnwidth]{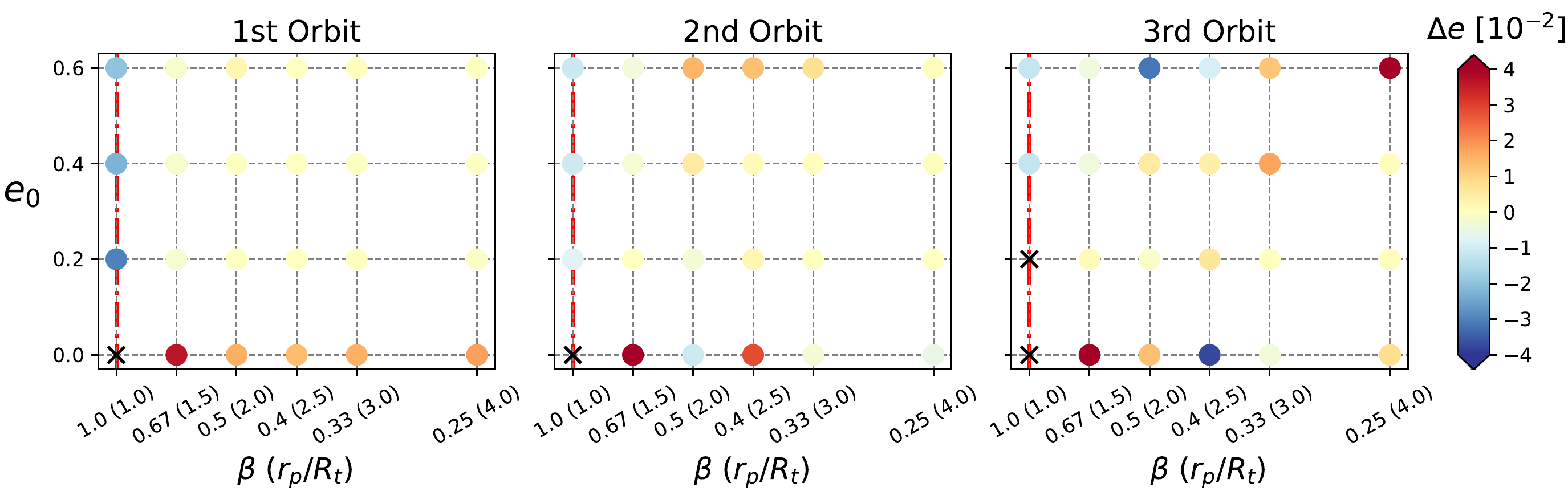}
    \caption{The change of eccentricity ($\Delta e$) during the first, second and third orbit around the BH.  The yellow colors represent (near) zero changes in eccentricity during the orbit, while the redder (bluer) points represent orbit becoming more (less) eccentric. }
    \label{fig:del_e}
\end{figure*}

Fig.~\ref{fig:del_e} shows the change of eccentricity $\Delta e$ in the first three orbits for the same models in Fig.~\ref{fig:del_a}. Most models show small variations in $\Delta e$ (yellow points), except for the initially circular models (bottom points) and the most compact models with different $e_0$ (points in the first column), which is consistent with the behaviors in $\Delta a$. The stars in these models are the most tidally influenced by the BH, where $\Delta e$ shows significant fluctuations in all three orbits -- some orbits become more eccentric then later circularize, and vice versa. 

\subsection{Sources of luminosity} \label{sec:luminosity}
In a TPE, the super-Eddington accretion onto the BH powers outflow from the accretion disk. 
% One caveat when predicting observed luminosity with the accretion rate from our simulation is that 
% $\dot{M}_{\rm BH}$ obtained from an SPH simulation can sometimes be overestimated by a few orders of magnitude. \cx{citation} Additionally, 
The EM emission from the TPE is delayed by the photon diffusion time ($\tau_{\rm diff}$), which dilutes the emission from the accretion disk. From our simulation, $\tau_{\rm diff} = \tau H / c \sim 10^5$ years, similar to the photon diffusion time in the sun. In this relation, $H\sim1.5R_{\odot}$ is the thickness of the accretion disk formed from the TPE. $\tau$ is the optical depth to electron scattering, computed assuming fully ionized gas as,
\begin{equation}
    \tau = \int_r^{\infty} \rho(r^{\prime}) \frac{\sigma_T}{m_p} dr^{\prime} \approx 10^{11},
\end{equation}
where $\sigma_T$ is the electron scattering cross-section. $\rho$ is the 3-dimensional density of the accretion disk taken directly from our simulations, which is typically very high since a large fraction of the star is stripped  to form the disk in a TPE.
% In reality, however, $\rho$ could be even larger due to the densest part of the stellar core that we do not recreate in our simulations.
Overall, the photon diffusion time $\tau_{\rm diff}$ is much longer compared to the viscous timescale of the accretion disk  \citep[eq. 4 in ][]{DOrazio2013},
\begin{align} \label{eq:t_visc}
    \tau_{\rm visc}& \simeq 1060 \left(\frac{\mathcal{M}}{10} \right)^{2}\left(\frac{0.01}{\alpha}\right)t_{\rm orb}\approx 20 {\rm \ days},
   %  \\
   % \tau_{\rm visc} &\simeq 20 ~{\rm days} \left(\frac{\mathcal{M}}{10} \right)^{2}\left(\frac{0.01}{\alpha}\right)\left(\frac{t_{\rm orb}}{0.5 {\rm hours}}\right),    
\end{align}
where $\mathcal{M}$ is the Mach number, $\alpha$ is the Shakura-Sunyaev viscosity parameter, and $t_{\rm orb}$ is the orbital time that is typically a few hours.
However, given the super-Eddington accretion rate of a TPE, a relativistic jet may be launched and
break out from the disk, possibly allowing the TPE to shine through. Since $\dot{M}_{\rm BH} \gg \dot{M}_{\rm Edd}$, there could be strong accretion disk outflow that might also modify the EM emission of a TPE. If the TPE is embedded in an AGN disk, the star and the BH will accrete mass from the disk. We use the calculations in \cite{Tagawa2020} to estimate that the mass accretion rates onto the star and the BH are both approximately $10^3\dot{M}_{\rm Edd}$, with the BH's accretion rate $\sim5$\% of the star's.  We assume that the TPE is located at $r\sim10^{-2}$ pc to the central massive BH of mass $10^6 M_{\odot}$, where the disk density  is $\rho_{\rm AGN}\sim10^{12}M_{\odot}/{\rm pc}^3$ and the aspect ratio is of $\simeq 10^{-3}$. The accretion rates from the AGN disk are also super-Eddington, although they are still few orders of magnitude lower than $\dot{M}_{\rm BH}$ in the TPE. Modeling these aspects of TPEs would require higher resolution, radiative transfer, and/or perhaps a different numerical code that can include the low-density background AGN disk, which could be addressed in future work. 

\section{Massive Stars}\label{sec:massive_stars}

\begin{figure*}
    \centering
    \includegraphics[width=\columnwidth]{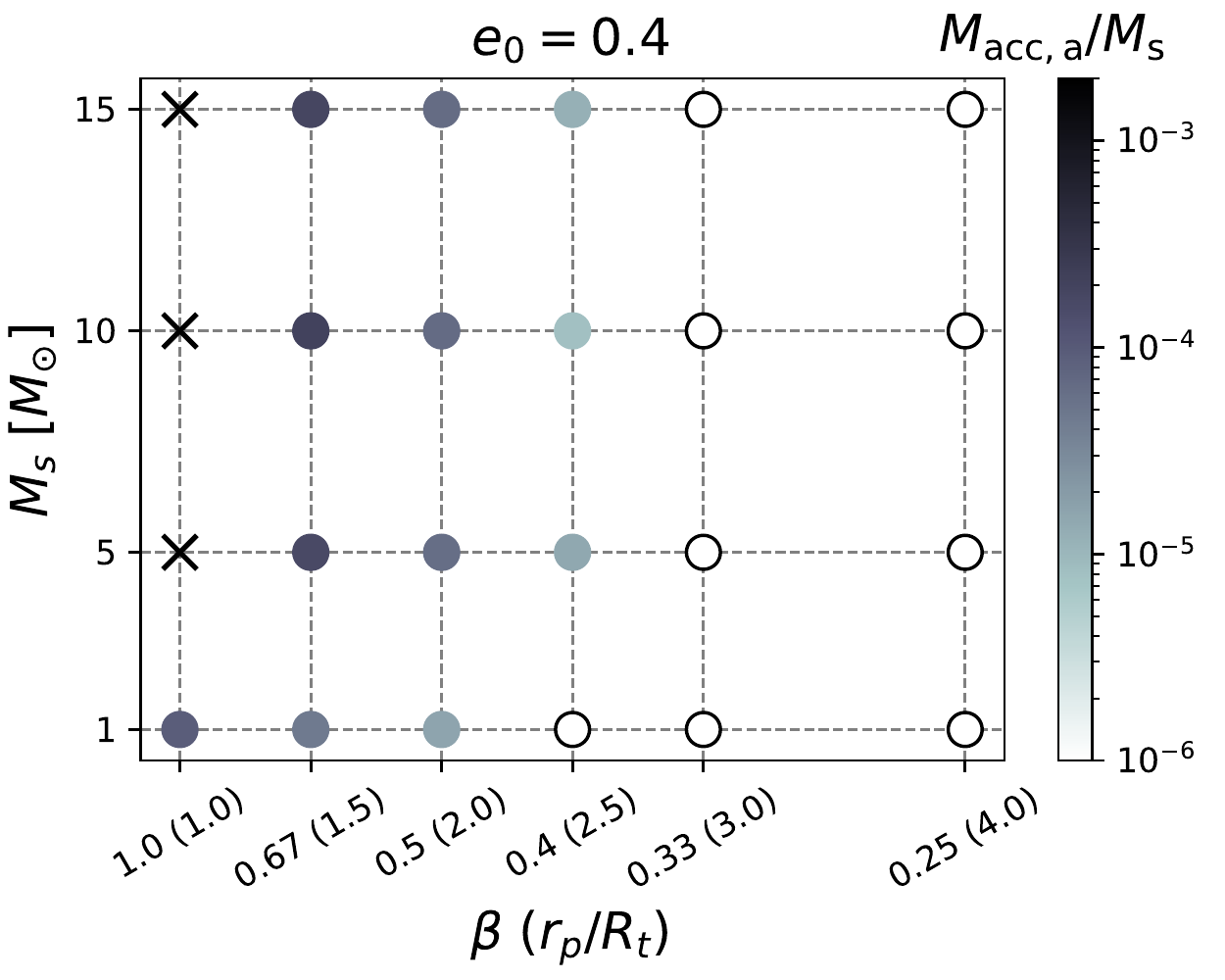}
    \includegraphics[width=\columnwidth]
    {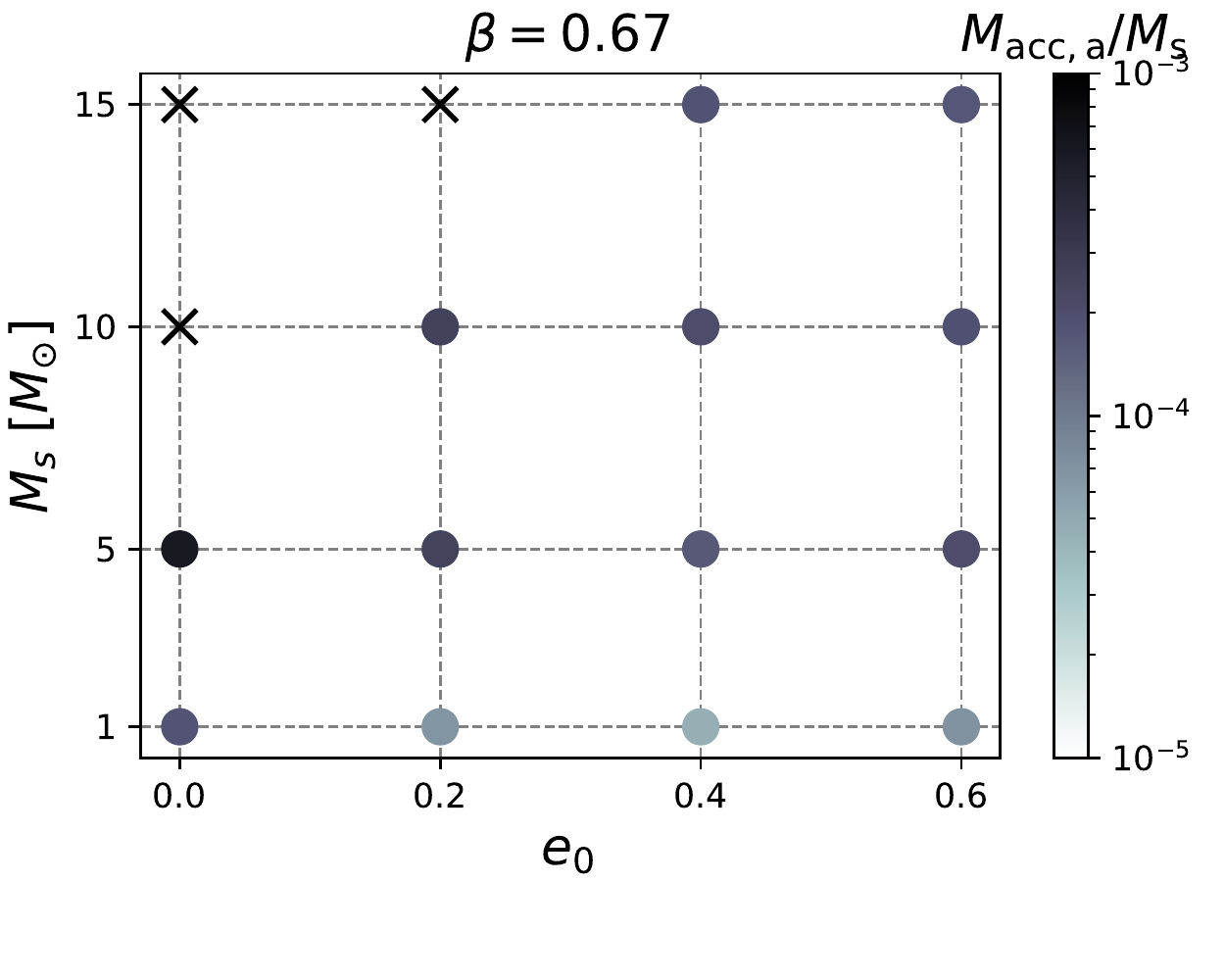}\\ 
    \includegraphics[width=\columnwidth]{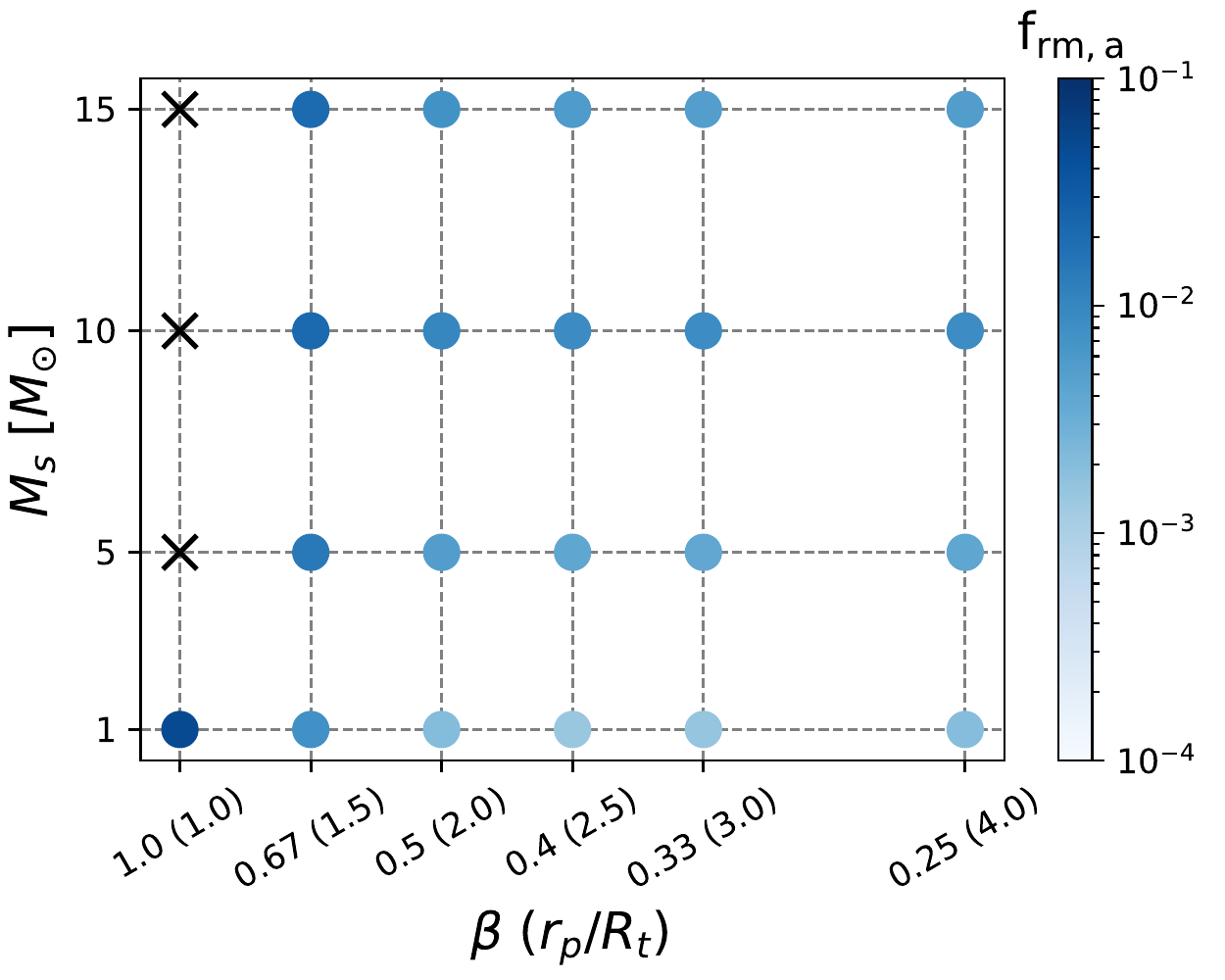}
    \includegraphics[width=\columnwidth]{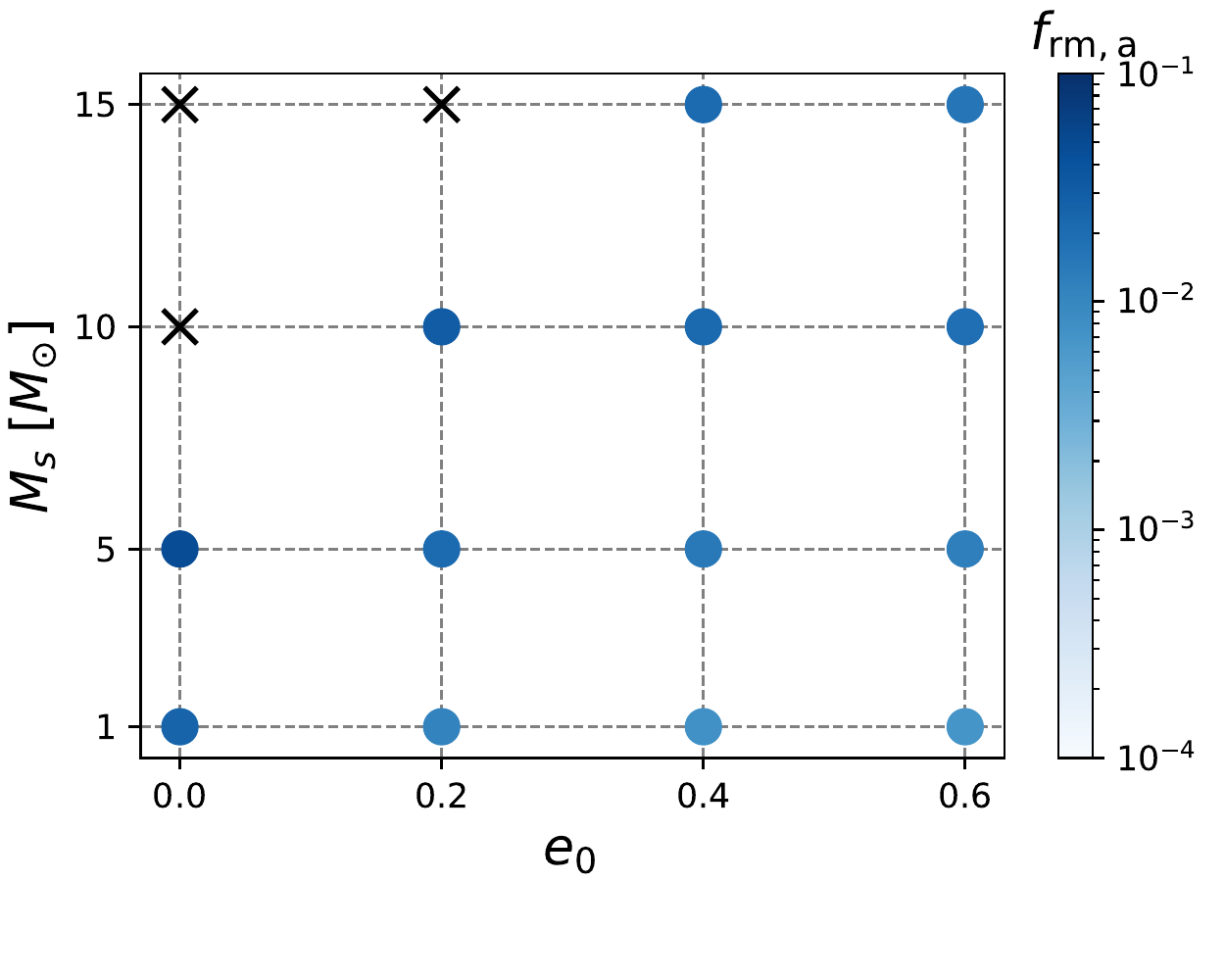}\\
    \includegraphics[width=\columnwidth]{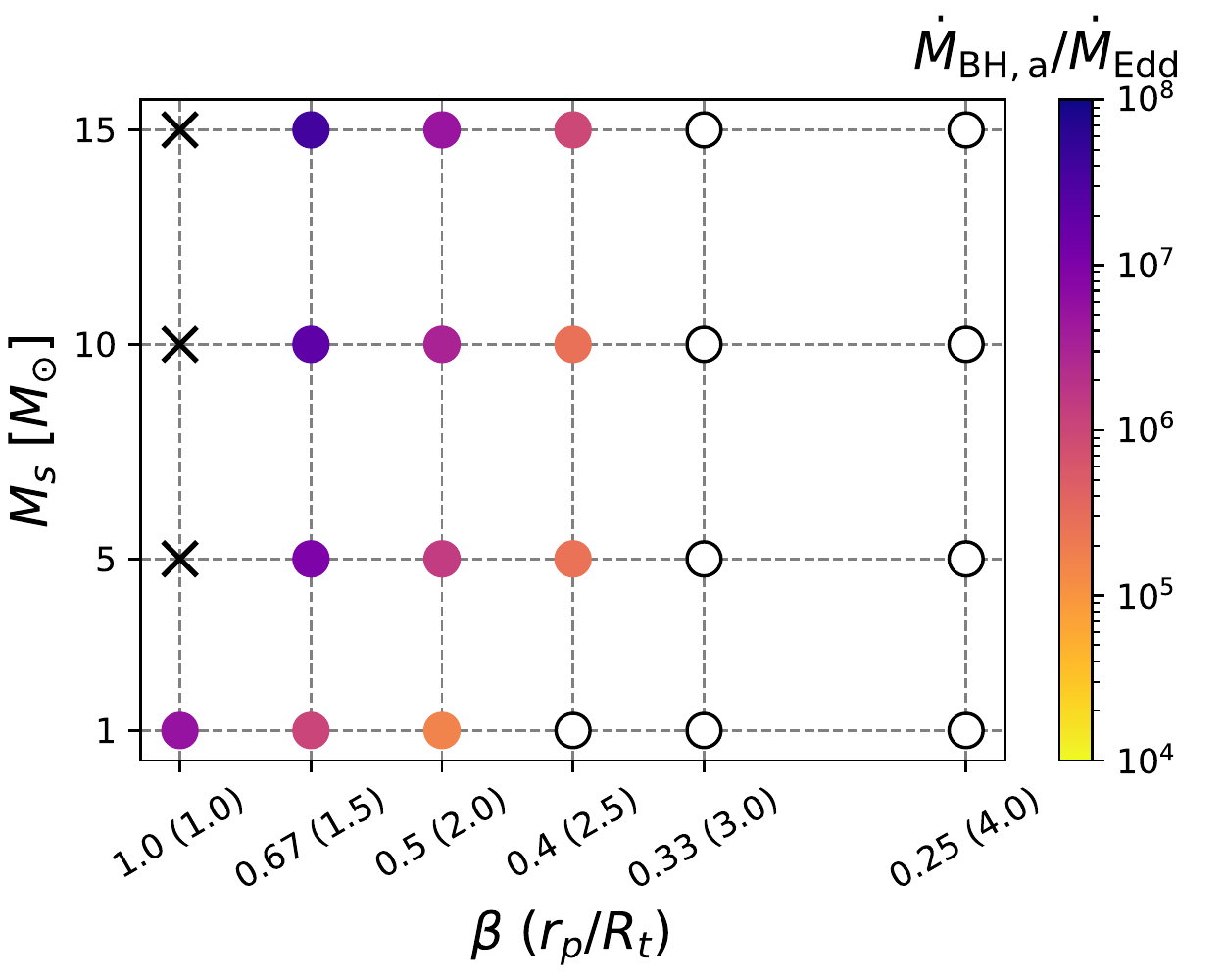} 
    \includegraphics[width=\columnwidth]{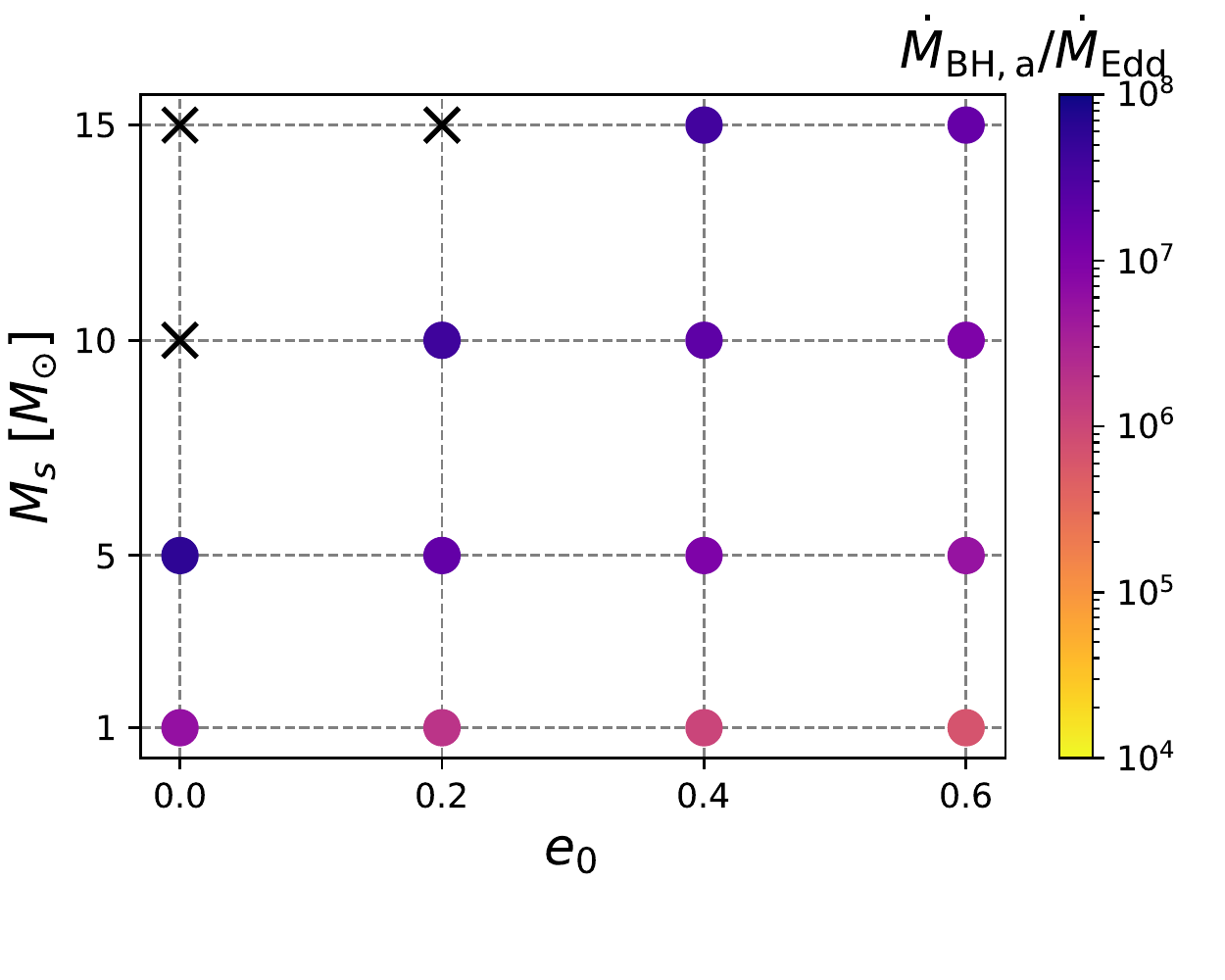}
    \caption{Mass accreted by the BH ($M_{\rm acc,a}$) normalized to stellar mass, the fraction of mass removed ($f_{\rm rm,a}$), and accretion rates onto the BH ($\dot{M}_{\rm BH,a}$) as function of (1) stellar mass and penetration factor at fixed initial $e_0$ = 0.4 (left column), and (2) stellar mass and eccentricity at fixed pericenter $\beta=0.67$. These are evaluated in the first orbit of the simulation. As in previous figures, cross indicates full disruption. In general, $M_{\rm acc,a}$, $f_{\rm rm,a}$ and $\dot{M}_{\rm BH,a}$ decreases for larger initial separation and eccentricity. The more massive stars are, the more likely to be completely disrupted due to larger stellar radius compared to tidal radius. There is a lack of trend in $M_{\rm acc,a}$, $f_{\rm rm,a}$ and $\dot{M}_{\rm BH,a}$ depending on $M_s$ on for $M_s >1M_{\odot}$, indicating that the stellar structures are not significantly different for those stars.}
    \label{fig:ecc_b1}
\end{figure*}

\begin{figure}
    \centering
    \includegraphics[width=\columnwidth]{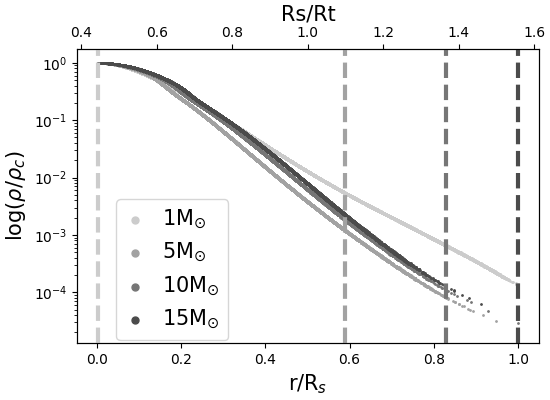}
    \caption{Initial density profiles of stars used in TPEs -- $M_s=1,5,10$ and $15M_{\odot}$ as labeled in the legend. The log-scale density (y-axis) is normalized by the core density of the star, which is a function of radius (bottom x-axis; normalized by stellar radius). The dashed line (top x-axis) of corresponding colors indicate the ratio of stellar radius to the tidal radius of each star.  }
    \label{fig:stars}
\end{figure}

Due to different stellar physics in massive stars, we investigate the behavior of TPEs where stars more massive than solar mass are involved, $M_s=5,10$, and $15M_{\odot}$.  
Fig.~\ref{fig:ecc_b1} shows (i) the properties of TPEs depending on the initial stellar mass and initial pericenter distance, at fixed $e_0=0.4$ (left panels), and (ii) the same properties depending on initial $M_s$ and $e_0$, at fixed $r_p$ (right panels). 
From top to bottom, we show the change in mass accretion onto the BH, the fraction of mass removed from the star, and mass accretion rate per orbit.
The crosses indicate that more massive stars are more likely to undergo total disruption given the same initial orbital configurations. In Fig.~\ref{fig:stars}, we see that this is because a more massive star's radius is closer to the pericenter, even though its density profile is steeper. Here we show the density profiles of the initial stars $M_s=1,5,10$ and $15M_{\odot}$, as labeled. The dashed lines (top x-axis) represent the ratio between stellar radius and tidal radius, which is larger for a more massive star. 

In Fig.~\ref{fig:ecc_b1}, we normalize the BH mass accretion by the initial mass of the star, $M_{\rm acc,a}/M_{\rm s}$ in the top two panels. Therefore, any change in $M_{\rm acc,a}/M_{\rm s}$, as well as in $f_{\rm rm,a}$ and $\dot{M}_{\rm BH}$, with the initial $M_s$ (along the y-axes) reflects different interior structures of the stars due to different masses. There are minimal changes in $M_{\rm acc,a}/M_{\rm s}$, $f_{\rm rm,a}$ and $\dot{M}_{\rm BH}$ along the $M_s$ axis, at any fixed $r_p$ or $e_0$, especially for $M_s\geq5M_{\odot}$. This indicates that the stellar interiors, mainly the envelopes that are responding to the tidal stripping of the BH, are not significantly different for different stellar masses, unless the core of the star is also disrupted, i.e. the cases of total disruptions. 

Overall, these three quantities show more variation due to different initial $r_p$ and $e_0$, compared to the effect of stellar mass. 
At fixed $e_0$, $M_{\rm acc,a}/M_{\rm s}$, $f_{\rm rm,a}$ and $\dot{M}_{\rm BH}$ decrease as the initial pericenter distance becomes wider, where $M_{\rm acc,a}/M_{\rm s}$ and  $\dot{M}_{\rm BH}$ reduce to zero (open circles) even for more massive stars. Similarly, at fixed $r_p$, these quantities decrease as $e_0$ gets larger, due to the fact that elliptical orbits with larger eccentricities (given the same pericenter distances) are longer orbits.  
Consistent with the $M_s=1M_{\odot}$ cases, the impact of $r_p$ is overall more significant than the impact of $e_0$. 
Generally, having a more massive star in the TPE results in more mass accretion onto the BH and higher accretion rates. Our figures show the fractions of star lost or accreted by the BH, which indicates the importance of different stars' interior structures. 

\begin{figure}
    \centering
    \includegraphics[width=\columnwidth]{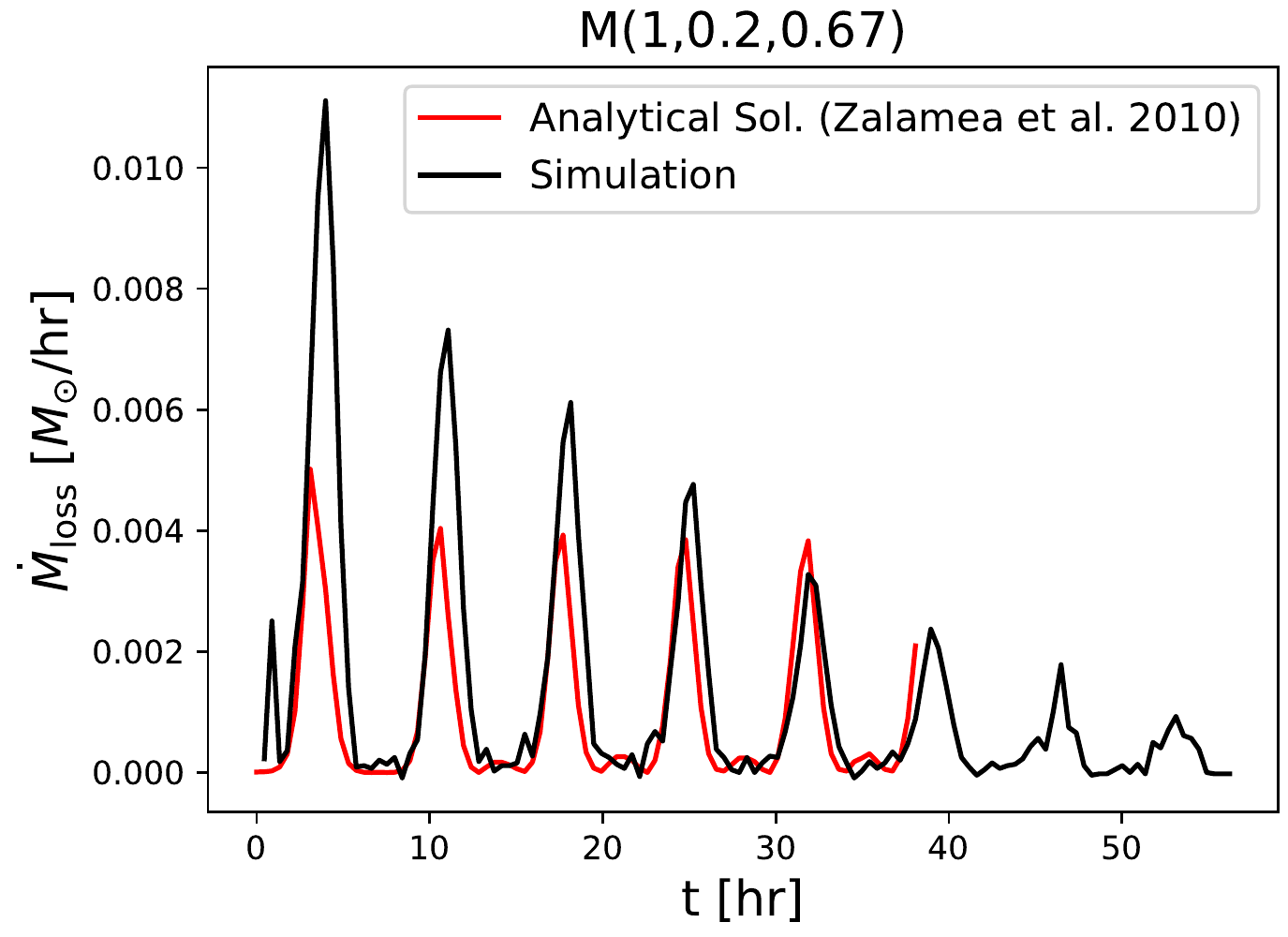}
    \caption{Comparing the rate of stellar mass loss from our simulation, $\mathcal{M}(1,0.2,0.67)$ (black), to that predicted by the analytical solution (red) in \citet{Zalamea2010}. }
    \label{fig:analy_predict}
\end{figure}

Finally, as a sanity check, we evaluate the mass loss rate of a $1M_{\odot}$ star from the analytical solution described in \citet{Zalamea2010}, and compare this solution to our simulation results. This analytical solution predicts the rate of mass loss of a white dwarf (WD) when it is tidally disrupted by a SMBH, which can be directly applied to our TPE scenario. \cite{Zalamea2010} predicts that an outer shell of the star with thickness $\Delta R$ is removed at each tidal stripping, as long as $R_s < 2R_t$, where $\Delta R = R_s - R_t \ll R_s$.
% , the mass of shell $\Delta R$ is lost at each pericenter, and the rate of mass loss is defined as $-\delta m/\tau(M)$ described in the following equations:
% \begin{equation}
%     \delta m = 4 \pi R_s^2 \int_0^{\Delta} \rho(z) dz
% \end{equation}
% and
% \begin{equation}
%     \tau(M) = (G \bar{\rho})^{-1/2}
% \end{equation}
% is the sound crossing timescale of the star.
The only differences being (i) our stellar density profile describes a solar-like MS star that is governed by gas+radiation pressure, instead of a WD governed by electron degeneracy pressure, and (ii) the pericenter is much closer to the tidal radius, since we have a stellar-mass BH rather than a SMBH. Adopting these changes, the analytical calculation of the stellar mass loss rate ($\dot{M}_{\rm loss}$; red) from our simulation is shown Fig.~\ref{fig:analy_predict}, along with the mass loss rate we evaluate from the simulation output (black). This figure shows reasonable consistency between the two, where the analytical solution is roughly half the simulation results at first. However, the analytical solution shows a slower drop in amplitude.

\section{Discussions}  \label{sec:discussion}
\subsection{Comparing TPEs to micro-TDEs, TDEs by intermediate-mass and supermassive BHs} \label{sec:comparison}
% \cx{compare period, luminosity, morphology, accretion rate, orbital evolution}
% 
\begin{figure}
    \centering
    \includegraphics[width=\columnwidth]{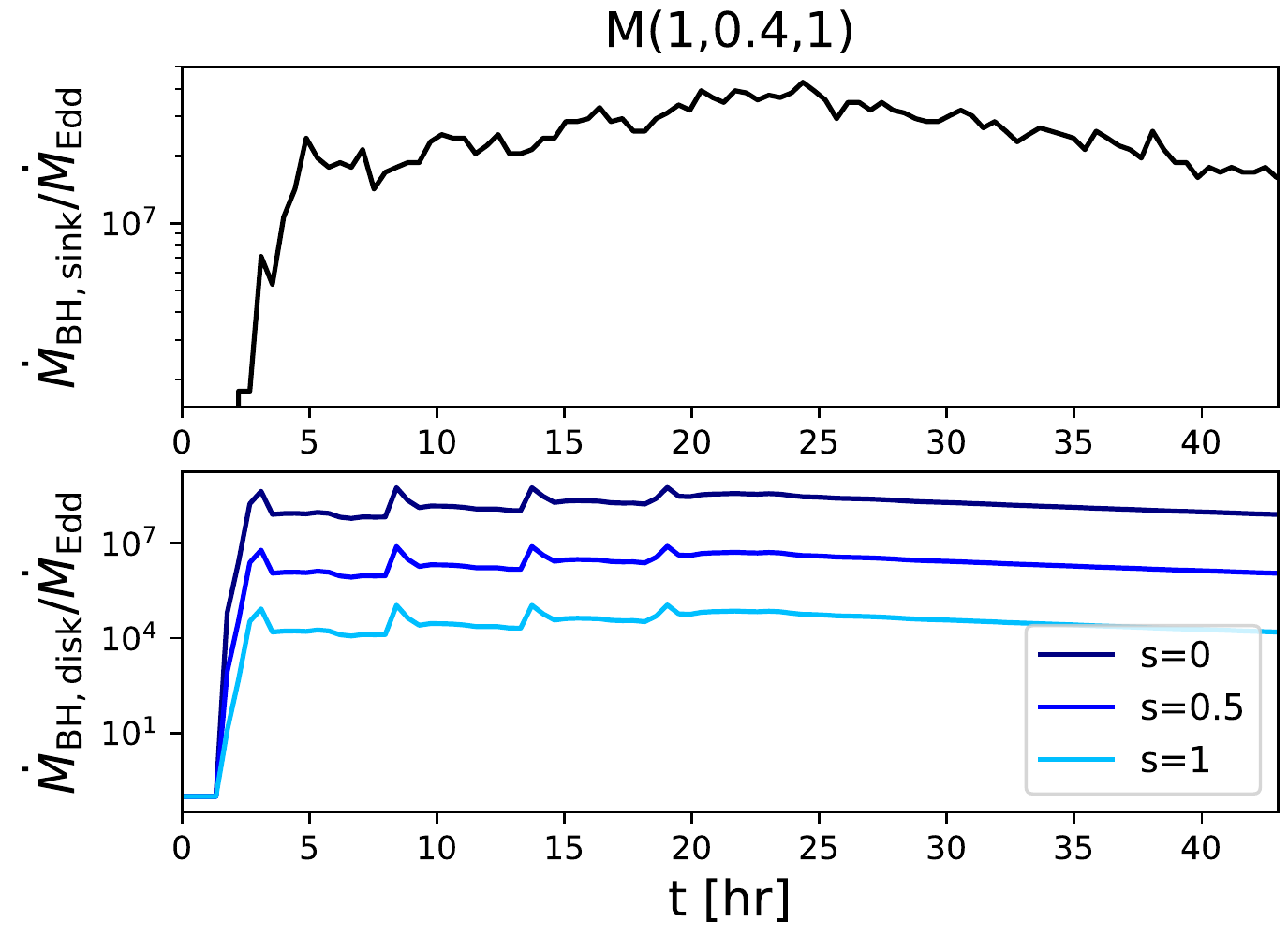}
    \caption{Accretion rates of the BH in TPE model $\mathcal{M}(1,0.4,1)$ evaluated using (i) mass accreted onto the sink particle (top panel) and (ii) using mass accretion calculation in eq.~\ref{eq:m_peak}, following \cite{Kremer2022} (bottom panel). The bottom panel adopts three choices of power-law index: $s=0,0.5$ and 1. }
    \label{fig:msink_mdisk}
\end{figure}
\begin{figure}
    \centering
    \includegraphics[width=\columnwidth]{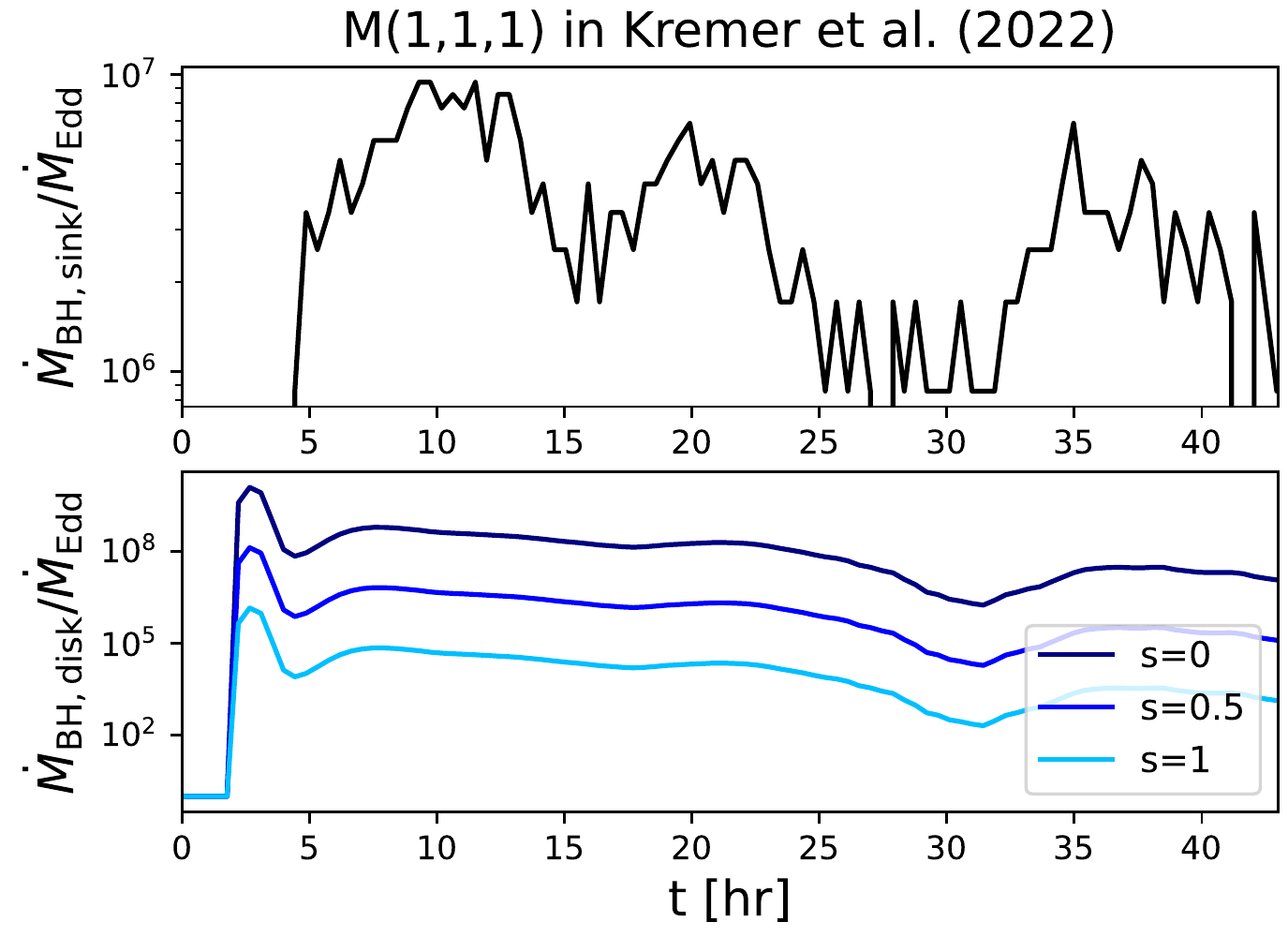}
    \caption{Same comparison as Fig.~\ref{fig:msink_mdisk}, but for parabolic micro-TDE model $\mathcal{M}(1,1,1)$ from \cite{Kremer2022}. Similar to Fig.~\ref{fig:msink_mdisk}, the accretion rates from the sink and disk methods show roughly consistent results. }
    \label{fig:msink_mdisk_k22}
\end{figure}
The orbit of the star in a micro-TDE is typically expected to be parabolic when tidally disrupted by the BH. From a recent study of hydro-simulations of micro-TDEs \citep[e.g.][]{Kremer2022}, they are likely ultra-luminous transients, similar to our finding for TPEs. We find that similar to micro-TDEs, TPEs have super-Eddington accretion rates, up to $\sim10^8\dot{M}_{\rm Edd}$, which is in order of magnitude comparable to that of ``normal" micro-TDEs, see Figure 11 in \cite{Kremer2022}. However, the method that \cite{Kremer2022} use to measure the accretion rate by assuming that some disk mass is accreted by the BH within the viscous time, or eq.~\ref{eq:m_peak}, is different from our method of using a sink particle to measure BH accretion rate. They assume:
\begin{equation} \label{eq:m_peak}
    \dot{M}_{\rm BH} \propto \Bigg(\frac{M_{\rm disk}}{t_{\rm visc}} \Bigg) \Bigg(\frac{R_{\rm in}}{R_{\rm disk}}\Bigg)^s.
\end{equation}
In this relation, we choose an accretion disk with radius $R_{\rm disk}=R_t=2.2R_{\odot}$ that includes particles within the initial Roche Lobe radius around the BH, see the last panel of Fig.~\ref{fig:morphology1}. $M_{\rm disk}$ is the disk mass, which eventually reaches $0.8M_s$. $t_{\rm visc}$ is the viscous timescale that we adopt from eq.~\ref{eq:t_visc}, but using Mach number $\mathcal{M}=3$ and $\alpha=0.1$. $R_{\rm in}$ is the inner edge of the disk -- we choose $R_{\rm in}=10 r_{\rm sch}$, where $r_{\rm sch}=2GM_{\rm BH}/c^2$. Finally, the choice of power-law index $s$ account for different levels of mass loss due to outflows. In Fig.~\ref{fig:msink_mdisk}, we compare the accretion rates computed with eq.~\ref{eq:m_peak} to that found with the sink particle, on a TPE model $\mathcal{M}(1,0.4,1)$. The top panel shows the mass accretion rate onto the sink particle ($\dot{M}_{\rm BH, sink}$), and the bottom panel shows the accretion rate from the disk calculation ($\dot{M}_{\rm BH, disk}$) assuming three choices for the power-law index $s=0,0.5,1$. $\dot{M}_{\rm BH, disk}$ is overall comparable to $\dot{M}_{\rm sink}$, while it rises earlier -- some mass falls within $R_{\rm disk}$ instantaneously after the simulation begins.  We perform the same comparison for a parabolic micro-TDE model that was reported in \cite{Kremer2022}, with $M_{\rm BH}=10M_{\odot}$, $M_{\rm s}=1M_{\odot}$, $e_0=1$ and $\beta=1$, see Fig.~\ref{fig:msink_mdisk_k22}. We adopt a disk with radius $R_{\rm disk}\sim 3.7R_{\odot}$, value used by \cite{Kremer2022}, and $t_{\rm visc}$ evaluated with Mach number $\mathcal{M}=1$. The two methods again yield similar accretion rates. 

Despite having similar accretion rates, the orbital periods are generally shorter for a TPE, which are between a few to few tens of hours, compared to periods of days to weeks for a micro-TDE. Some micro-TDE models in \cite{Kremer2022}, such as the model with a more massive $M_s=5M_{\odot}$ star and a $10M_{\odot}$ BH, show multiple passages and therefore periodic accretion onto the BH just like in TPE. However, the orbital period in this model is $\sim$4 days, significantly longer than TPE periods, so we will be able to distinguish that from a TPE. But generally, stars in most micro-TDEs undergo tidal stripping only once, leaving very different morphological evolution, accretion and orbital signatures compared to a TPE. 
% Compare tidal peeling events to the more typical micro-TDE with $e=1$ and TDE with massive BHs. Here we will compare the similarities and differences between our TPE simulations to a recent study that performed hydro simulations of micro-TDEs, \citet{Kremer2022} and other works e.g. \citet{Kroglu2022}. Peak luminocity plots. 

Another important comparison should be made between TPEs and tidal disruptions of a solar-like star by an intermediate-mass BH (IMBH). Recent work by \cite{Kroglu2022} find, using hydro-simulations, that in all cases where a 1$M_{\odot}$ star is disrupted by a IMBH, the stellar remnant is eventually ejected to be unbound, either after the first pericenter or after many pericenter passages. In our TPE simulations, all stars remain in a binary with the BH, or are eventually completely disrupted by the BH. If the star survives for many pericenter passages with IMBH, then the star is only partially disrupted and the accretion rate increases with the number of orbits. This is also not the case in TPEs, see the RHS of Fig.~\ref{fig:6-panel1}, where $\dot{M}_{\rm BH}$ decreases with the number of orbits.  Finally, the orbital periods of tidal disruptions by IMBH typically span a wide range, from 10s of hours to 10 thousand years. The shortest-period events with comparable periods to TPEs correspond to the lowest BH mass ($M_{\rm BH}\lesssim$10$M_{\odot}$) and smallest pericenter distance ($r_p/R_t\ll1$). Therefore, these events are basically the micro-TPEs in \cite{Kremer2022} and their similarities and differences to TPEs are already mentioned above. 

The best indicator that a micro-TDE is present in an AGN, rather than a TDE of a solar-like star by a SMBH, is if the mass of the SMBH is above the Hill's limit $\gtrsim10^8 M_{\odot}$, beyond which the Schwarzschild radius of the BH is greater than the tidal radius. However, micro-TDEs or TPEs have distinguishable signatures even if they exist near a smaller SMBH. First, the spectra of micro-TDEs and TDEs are expected to be very different, because the remnant produced in micro-TDEs tend to be optically thick -- this is even more so the case in TPEs, which lead to a hotter accretion disk that cools less efficiently \citep{Wang2021} and result in emission in the higher-end of X-rays. Additionally, like in most micro-TDEs, SMBH in a TDE typically will disrupt the star once and strip $\sim$half to all of its mass, while partial disruptions are more common in TPEs. Partial disruptions in TDEs, however, will have periodic flares on a yearly scale, such as recently observation of repeated bursts in AT2018fyk \citep{Wevers2022}, much longer than the expected periods of micro-TDEs and TPEs. 

Overall, our simulations show that TPEs are novel transient phenomena that can be distinguished from other ultra-luminous transients such as micro-TDEs, tidal disruptions by IMBHs and SMBHs, and partial disruptions in sTDEs. 
% Theoretical investigations of TPEs have many important implications, such as understanding interactions in compact star-BH binaries in star clusters or AGN disks and observations of ultra-luminous transient events especially those near the galactic center. 

\subsection{Simulation caveats} \label{sec:caveats}
Theoretical investigations of TPEs have many important implications, such as understanding interactions in compact star-BH binaries in star clusters or AGN disks and observations of ultra-luminous transient events especially those near the galactic center. 
Our results offer first-hand understanding of TPEs with simulations, while they should be treated as numerical experiments rather than accurate physical descriptions of TPEs in a cluster or embedded in an AGN disk. We list the following caveats of our simulations that should be improved in the future. First, we start the simulations with already very compact orbits, while in reality, they should be expected at the end of some dynamical process such as a long AGN-disk mediated inspiral or interactions between multiple stars or compact objects in a star cluster.  Since the BH and star should approach each other from a much larger distance, we might expect the star to have already been partially disrupted by the BH, although no mass will be accreted by the BH beyond the separation of $r_p\sim3R_t$, as shown in our results. The binary could, however, accrete from the external AGN gas, if embedded in an AGN disk.
In future work, we will investigate the effect of torques from the circumbinary gas on binary, which can shrink the orbital separation.
Additionally, one could also include the low-density AGN disk gas as a background of the TPE simulations, instead of using vacuum. This is challenging with SPH simulations, but could instead be feasible with grid-based codes. Finally, it is also important to add radiation outflows from the optically-thick accretion disk and shock properties due to the relative motion of the star/BH and the debris, in order to more accurately describe TPEs. 
Future work should perform simulations or make analytical predictions for TPEs considering all of these additional factors above.

\subsection{Detectability of TPEs as transients in AGNs} \label{sec:detectability}

The AGNs are extremely dynamical locations to host luminous transients. Identifying TPEs among different transient events in AGNs will require careful examination of their EM signatures. 
% In additional to the comparison we make between TPEs, micro-TDEs, tidal disruption by IMBHs and partial TDEs in \S~\ref{sec:comparison}, the distinction also need to be made for tidal disruptions by/of other compact objects, such as the tidal disruption of a white dwarf \citep{Maguire2020} and neutron star by a stellar-mass BH, and tidal disruption of a star by a neutron star. 
AGNs around heavy SMBHs ($M_{\rm SMBH} \gtrsim 10^8 M_{\odot}$) are shown to be the ideal place for identifying micro-TDEs \citep{Yang2022} and other transients alike. In order to observe TPEs, they need to outshine the AGN disk. Since our results show that TPEs result in super-Eddington accretion onto the BH, there could be super-luminous jet launching from the BH. Therefore, the EM emissions from TPEs can be subject to jet modulation, among many other mechanism such as accretion disk outflows and shocks, as mentioned in \S~\ref{sec:luminosity}. Even though the accretion disk formed from the stellar remnant is optically thick, and the AGN can also trap the radiation, the emissions from TPEs can be more visible if (i) the jet can eject gas from the circumbinary disk \citep{Tagawa2022}, and (ii) stellar-mass BHs can open cavities in the AGN disk \citep{Kimura2021} -- both of these will reduce the opacity of the surrounding gas. 
Finally, if the AGN does not launch any jets, then TPEs can outshine the AGN more easily in the radio or in the gamma rays. 

Here, we focus on the existing observational signatures of two micro-TDE candidates observed in AGNs that might also indicate TPE origins. 
Micro-TDE candidates in AGNs with a SMBH too massive for tidal disruption of a solar-type star \citep[ASASSN-15lh and ZTF19aailpwl;][]{Yang2022}, have peak luminosity $L_{\rm peak}\approx 5\times 10^{45}$ erg s$^{-1}$ and $L_{\rm peak}\approx 10^{45}$ erg s$^{-1}$. \cite{Yang2022} hypothesize that the higher peak luminosity of ASASSN-15lh indicates a micro-TDE, unless it is a result of tidal disruption of a star more massive than solar. From our simulations, we see that TPEs with a more massive star also produce higher accretion rates. The observations of ZTF19aailpwl show a longer rise time than a typical TDE, indicating a more gradual tidal disruption than a TDE with SMBH, e.g. produced by micro-TDEs with low eccentricity such as a tidal peeling event. 
% There have been some discussions on the possible observational features of micro-TDEs \citep[e.g][]{Bartos2017,Yang2022,Wang2022}. 
Finally, the rate of micro-TDEs are expected to be low in AGNs, at roughly $2$ Gpc$^{-3}$ yr$^{-1}$ \citep{Yang2022}, and even lower in star clusters or stellar triple systems with BHs, while these predictions have large uncertainties. Only the brightest events are expected to be eventually observed, since the emission of most weaker micro-TDEs and TPEs will be dimmed significantly by the surrounding AGN gas. The mechanism that the emission from an event like the TPE propagates through an AGN disk is analogous to the propagation of GRB afterglow in a dense medium \citep{Perna2021,Wang2022}. Therefore, bright TPEs might have observational signatures similar to that of ultra-long GRBs. 

% \subsection{Stability of binaries in AGN disks} Wu and Quataet 2022 calculaion

\section{Summary} \label{sec:summary}
In this paper, we perform the first hydro-simulations of TPEs with the SPH simulation code \texttt{PHANTOM} to investigate their morphology, accretion signature and orbital evolution. We explore a range of initial conditions, including stellar mass, initial eccentricity and penetration factor, which make up 96 simulation models in total. We examine the impacts of these initial parameters on the behaviors of TPEs. 

First, we observe the ``tidal peeling" feature from our simulations where a solar-like or massive star is slowly and periodically tidally disrupted by a stellar-mass BH and its mass is slowly removed over many orbits. 
Due to low eccentricity, the orbital periods of TPEs are generally shorter ($P\sim{\rm few}$-few 10s of hours) compared to the micro-TDEs and TDEs.
In the most compact orbits, $r_p\approx R_t$, the star gets completely disrupted very quickly, after $\sim$1-4 orbits; otherwise, the star ends up being partially disrupted. 
Out of the three initial conditions, the penetration factor has the largest effect on the accretion and orbital signatures of interest, namely mass accreted onto the BH, accretion rate, the fraction of mass removed from the star, the orbital separation, semi-major axis and eccentricity. As the orbit becomes more compact, there is more mass accreted by the BH, higher accretion rate and higher fraction of mass removed from the star. Lower eccentricity has a similar effect, since lower $e_0$ means that the orbit is shorter (recall that the star is placed at the apocenter at the start of the simulations). A few models with higher eccentricities show a periodic fluctuation in $\dot{M}_{\rm BH}$ that peaks after each pericenter passage. 

The orbital separation, semi-major axis and eccentricity demonstrate less obvious trends, especially when $\beta<1$ (less compact systems).  It is clear from the fluctuations in $a$ and $e$ that the orbit of a star in a TPE deviates from Keplerian due to the tidal influence and possibly also shocks from the stellar remnant encountering the tidal streams. In the most compact configurations, $\beta=1$, the orbital separation always shrinks regardless of the choice of $e_0$ and $M_s$, so both the semi-major and eccentricity decrease with the number of orbits. In these cases, the star is always completely disrupted at the end, consistent with the analytical limit of onset mass loss of tidal stripping at $\beta=1$ \citep[e.g.][]{Zalamea2010}. 
Finally, if there is a more massive star in the TPE, the stellar radius is larger and, at fixed $\beta$, it is closer to the tidal radius. Therefore, the disruption is more rapid and total disruption of the star is more common. There is higher mass loss from the star as well as more accretion by the BH. However, for stars more massive than $1M_{\odot}$, the fraction of the initial stellar mass lost or accreted by the BH does not vary significantly due to different stellar masses. This indicates the similarity in the stellar structures of the more massive stars. 

The resulting accretion rates of TPEs are typically highly super-Eddington, $\dot{M}_{\rm BH}\sim10^{4-8}\dot{M}_{\rm Edd}$. However, since the accretion disk formed from the dense stellar material around of the BH is extremely opaque, the emission from TPEs will be affected by photon diffusion. Other mechanisms might exist to modulate the luminosity of the TPE, other than the BH accretion rate, such as relativistic jet launching from the BH and shocks due to relative motion of the star remnant and the tidal streams. A jet might empty a cocoon of low-density region around the TPE, possibly allowing the emission to be less affected by the thick accretion disk or AGN disk. 
Our results are also subject to a few caveats due to the limitations of our simulations. Future work should address more realistic aspects of TPEs, such as the radiation for the hot accretion disk, shocks, binary inspiral from a farther separation, and/or AGN background gas. 

Finally, better theoretical understanding of TPEs is highly motivated by the existing observations of abnormal flaring events from AGNs, such as SASSN-15lh and ZTF19aailpwl, that can not be well explained by AGN variability, or other luminous transients such as TDEs by SMBHs. AGNs are extremely dynamical playgrounds for interacting stars and compact objects. Our results suggest that identifying TPEs among many different ultra-luminous transients can be feasible due to its unique accretion signatures and orbital evolution that we find in this work.

\section*{Acknowledgements}
ZH acknowledges support from NASA grant 80NSSC22K082. 
RP acknowledges support by NSF award AST-2006839. YW acknowledges support from Nevada Center for Astrophysics. CX acknowledges the support from the Department of Astronomy 
 at Columbia University for providing computational resources for this research. 

% \software{
%     Astropy \citep{Astropy2013, Astropy2018},
%     Hasasia \citep{HazbounHas:2019},
%     Healpy \citep{Healpy2019}, 
%     Matplotlib \citep{Hunter2007}, 
%     Numpy \citep{Numpy2011},
%     Pandas \citep{Pandas2010}, 
%     SciPy \citep{SciPy2019},
%     NanohertzGWs \citep{MingarelliGWs}
% }

\bibliography{cx}

\end{document}